\documentclass[%
 reprint,
nofootinbib,
 amsmath,amssymb,
 aps,
]{revtex4-1}

\usepackage{graphicx}
\usepackage{dcolumn}
\usepackage{bm}
\usepackage{amsmath,amsfonts,amssymb,slashed,upgreek,color}
\usepackage{multirow}
\usepackage{xcolor}
\usepackage{hyperref}
\usepackage{xspace}
\graphicspath{{plots/}}

\newcommand\ee{\end{equation}}
\newcommand\be{\begin{equation}}
\newcommand{\zb}{z_b}
\newcommand{\zf}{z_f}
\newcommand{\bn}{\mathbf{n}}
\newcommand{\deltaim}{\Delta_{\rm HI}}
\newcommand{\ec}{E^{\times}_\ell}
\newcommand{\ecl}{E^{\times{\rm len}}_\ell}
\newcommand{\ecc}{E^{\times{\rm c}}_\ell}
\newcommand{\hec}{\hat{E}_\ell^{\times}}
\newcommand{\es}{E_\ell^{\rm st}}
\newcommand{\esl}{E_\ell^{\rm st\,len}}
\newcommand{\esc}{E_\ell^{\rm st\,c}}
\newcommand{\hes}{\hat{E}_\ell^{\rm st}}
\newcommand{\im}{{\rm HI}}
\newcommand{\FF}{\mathcal{F}}
\newcommand{\mnras}{Mon. Not. Roy. Astron. Soc.}

\newcommand{\newl}{GIMCO\xspace}

\newcolumntype{C}[1]{>{\centering\arraybackslash}p{#1}}
\newcolumntype{L}[1]{>{\raggedright\arraybackslash}p{#1}}
\newcolumntype{R}[1]{>{\raggedleft\arraybackslash}p{#1}}

\begin{document}


\title{A new estimator for gravitational lensing using galaxy and intensity mapping surveys}

\author{Mona Jalilvand}
\email{Mona.Jalilvand@unige.ch}
\author{Elisabetta Majerotto}
\author{Camille Bonvin}
\author{Fabien Lacasa}
\author{Martin Kunz}
\affiliation{%
Universit\'e de Gen\`eve, D\'epartement de Physique Th\'eorique and Centre for Astroparticle Physics, 24 quai Ernest-Ansermet, CH-1211 Gen\`eve 4, Switzerland
}%
\author{Warren Naidoo}
\author{Kavilan Moodley}
\affiliation{%
Astrophysics \& Cosmology Research Unit, School of Mathematics, Statistics \& Computer Science, University of KwaZulu-Natal, Westville Campus, Durban 4000, South Africa
}%

\date{\today}

\begin{abstract}
We introduce the Galaxy Intensity Mapping cross-COrrelation estimator (GIMCO), which is a new tomographic estimator for the gravitational lensing potential, based on a combination of intensity mapping (IM) and galaxy number counts. The estimator can be written schematically as IM$(z_f)\times$galaxy$(z_b)$ $-$ galaxy$(z_f)\times$IM$(z_b)$ for a pair of distinct redshifts $(z_f,z_b)$; this combination allows to greatly reduce the contamination by density-density correlations, thus isolating the lensing signal. As an estimator constructed only from cross-correlations, it is additionally less susceptible to systematic effects. We show that the new estimator strongly suppresses cosmic variance and consequently improves the signal-to-noise ratio (SNR) for the detection of lensing, especially on linear scales and intermediate redshifts. 
For cosmic variance dominated surveys, the SNR of our estimator is a factor 30 larger than the SNR obtained from the correlation of galaxy number counts only.
Shot noise and interferometer noise reduce the SNR. For the specific example of the Dark Energy Survey (DES) cross-correlated with the Hydrogen Intensity mapping and Real time Analysis eXperiment (HIRAX), the SNR is around 4, whereas for Euclid cross-correlated with HIRAX it reaches 52. This corresponds to an improvement of a factor 4-5 compared to the SNR from DES alone. For Euclid cross-correlated with HIRAX the improvement with respect to Euclid alone strongly depends on the redshift. We find that the improvement is particularly important for redshifts below 1.6, where it reaches a factor of 5. This makes our estimator especially valuable to test dark energy and modified gravity, that are expected to leave an impact at low and intermediate redshifts. 
\end{abstract}

\pacs{Valid PACS appear here}
\maketitle


\emph{Introduction.}
Gravitational lensing is a powerful probe of the matter distribution in our Universe. It describes the deflection of light rays by metric perturbations along the photon trajectory from their distant sources. Weak gravitational lensing refers to the regime where the deflections are small enough to not induce caustics.
The most common approach to observe weak lensing is through the distortion of  the observed shape of galaxies, which generates correlations between their ellipticity. This effect, referred to as cosmic shear, has been detected for the first time in the early 2000s~\cite{Bacon:2000sy,Wittman:2000tc,vanWaerbeke:2000rm,Kaiser:2000if}, and has been subsequently measured in various surveys providing tests of the consistency of the $\Lambda$CDM model~\cite{Troxel:2017xyo,Hildebrandt:2018yau,Hamana:2019etx, Jee:2015jta}. But weak lensing also modifies the observed number of distant galaxies, via the effect of lensing magnification (also called  magnification bias):
weak lensing modifies on one hand the observed size of the solid angle in which we count how many galaxies we detect, consequently diluting the number of galaxies per unit of solid angle. On the other hand, weak lensing modifies the observed luminosity of galaxies, enhancing consequently the number of galaxies that are above the magnitude threshold of a given survey. These two effects combine to distort the number counts of galaxies~\cite{Scranton:2005ci,Duncan:2013haa}.

One challenge in measuring cosmic shear
comes from the fact that it requires precise images of galaxies. Lensing magnification has the advantage of not relying on precise imaging, since the effect is measured from the galaxy number counts. However, it is affected by intrinsic fluctuations in the number counts of galaxies, which generically strongly dominate over {lensing magnification}~\cite{schneider}. This can be overcome by correlating galaxies at widely separated redshifts. In this case, density fluctuations become uncorrelated, and the only remaining signal of {cosmological interest} comes from lensing. {Lensing magnification} has been robustly measured using this technique: for example \cite{Scranton:2005ci} has measured the cross-correlation of quasars at redshift $1<z<2.2$, with foreground galaxies at mean redshift 0.3 in SDSS; whereas \cite{Garcia-Fernandez:2016oud} has used the cross-correlation of background galaxies at redshift $0.7 < z < 1$ with foreground galaxies at $0.2 < z< 0.4$ in DES. 

A new approach to map the large-scale structure of the Universe up to high redshifts is 
intensity mapping with radio telescopes.
These surveys will observe the intensity fluctuations of some emission line, typically  the 21\,cm line emitted by neutral hydrogen that is expected to trace the fluctuations in the galaxy distribution \cite{Pritchard:2011xb}. Various existing or planned post-reionisation radio surveys, like BINGO, HIRAX, CHIME, MeerKLASS, Tianlai, {PUMA,} and the SKA \cite{2013MNRAS.434.1239B, Newburgh:2016mwi, Newburgh:2014toa, Santos:2017qgq, Xu:2014bya, Bandura:2019uvb, Bull:2015lja} will detect these fluctuations and measure the power spectrum of the 21\,cm radiation. 

In this letter we propose a novel method to measure lensing magnification by correlating the fluctuations in 21\,cm intensity mapping (or intensity mapping of other lines) with the galaxy number counts, in such a way as to isolate lensing magnification. The main idea is that lensing magnification affects the galaxy number counts, but has no impact on intensity mapping. Due to the conservation of surface brightness, the number of photons that are lensed into our solid angle of observation by gravitational lensing is exactly compensated by the apparent increase of this solid angle~\cite{Hall:2012wd} ~\footnote{This is not the case at second-order in perturbation theory, where fluctuations in the 21\,cm intensity are themselves lensed~\cite{Jalivand:2018vfz}.}. As a consequence, the following schematic estimator, named \newl, isolates the lensing magnification contribution 
\be
\label{eq:est}
{\rm IM}(\zf) \times {\rm galaxy}(\zb)-{\rm galaxy}(\zf) \times  {\rm IM}(\zb)\, ,
\ee
where $\zb$ refers to the background redshift, and $\zf<\zb$  is the foreground redshift. In the first term of Eq.~\eqref{eq:est}, galaxies in the background are lensed by the presence of foreground matter perturbations, responsible for the 21\,cm signal. In the second term, however, the 21\,cm intensity in the background is not lensed.
By subtracting the two terms, we cancel the density-density correlations that affect both terms in the same way (up to bias differences), while keeping the lensing magnification contribution. This method, therefore, provides a way to \emph{isolate} lensing magnification, without restricting ourselves to wide redshifts separations. In the next section, we elaborate on this idea, and we show how the \newl estimator increases the signal-to-noise of lensing magnification by a factor of $\sim$ 30 for cosmic variance dominated surveys, and a factor of 4-5 for specific examples like DES$\times$HIRAX and Euclid$\times$HIRAX.

\vspace{0.2cm}

\emph{Estimator.} The galaxy number counts in direction $\bn$ and redshift $z$ are given by~{\cite{schneider}}
\begin{align}
\label{eq:Deltag}
\Delta_g&(\bn, z)= b_g(z) \, \delta(\bn, z)+\left(2-5s(z)\right) \, \phi(\bn, z)\, ,
\end{align}
where $b_g$ is the galaxy bias, $s$ is the slope of the luminosity function and $\delta$ denotes the matter density fluctuations. {Lensing magnification} (also called magnification bias), is the second contribution, proportional to the lensing potential \be
\label{eq:phi}
\phi(\bn, z)=-\int_{0}^{r} dr' \,\frac{r -r' }{rr'} \, \Delta_{\Omega} (\Phi+\Psi)\, ,
\ee
with  $r$ the conformal distance to the source, $\Phi$ and $\Psi$ the two metric potentials, and $\Delta_{\Omega}$ the angular part of the Laplacian. 

Intensity mapping is generically expressed in terms of the brightness temperature, whose fluctuations are a biased tracer of matter density
\be
\label{eq:DeltaHI}
\deltaim(\bn, z)= b_{\rm HI}(z) \ \delta(\bn, z)\,,
\ee
where $b_{\rm HI}$ is the bias of neutral hydrogen.
We neglect in Eq.~\eqref{eq:Deltag} and~\eqref{eq:DeltaHI} the contribution from redshift space distortions since we will average our estimator over thick redshift bins {of size $\Delta z=0.1$}. 
We also neglect the contribution from relativistic effects~\cite{Yoo:2009au,Bonvin:2011bg,Challinor:2011bk,Jeong:2011as,Hall:2012wd} which are subdominant in the regime we are interested in. 

We can expand the number counts and the brightness temperature fluctuations in spherical harmonics
\be
\Delta_X(\bn, z)=\sum_{\ell m}a^X_{\ell m}(z) \ Y_{\ell m}(\bn)\, ,
\ee
with $X=g, {\rm HI}$.
We now define {the \newl} estimator which cross-correlates galaxies and 21\,cm intensity mapping:
\begin{eqnarray}
\label{eq:Ecross}
\hec & \equiv& 
\hat{C}_\ell^{{\rm HI} g}(\zf,\zb) - \hat{C}_\ell^{g {\rm HI}}(\zf,\zb) \\
& = & \frac{1}{2\ell +1}\!\sum_{m=-\ell}^{\ell}\! \Big[a^{*{\rm HI}}_{\ell m}(\zf)a^g_{\ell m}(\zb)-a^{*g}_{\ell m}(\zf)a^{{\rm HI}}_{\ell m}(\zb)\Big]\, . \nonumber
\end{eqnarray}
Here $\hat{C}_\ell$ is a general estimator for the angular power spectrum, and the second equality holds for the standard full-sky estimator.
The expectation value of $\hec$ is
\begin{align}
\label{eq:meancross}
\ec & \equiv \langle\hec \rangle = 
\frac{1}{2}b_{{\rm HI}}(\zf) (2-5s(\zb)) C^{\delta\phi}_{\ell} (\zf,\zb) \\
&- \frac{1}{2} b_{{\rm HI}}(\zb)(2-5s(\zf)) C^{\phi\delta}_{\ell} (\zf, \zb) \nonumber \\
&+\big[b_{{\rm HI}}(\zf) b_g(\zb) - b_g(\zf) b_{{\rm HI}}(\zb)\big] C^{\delta \delta}_\ell(\zf, \zb)\,. \nonumber 
\end{align}
The first line is the contribution we want to measure: it represents the lensing potential of background galaxies generated by a foreground density at $z_f$. The second line, which contains the correlation between the lensing potential in the foreground and the density in the background, is negligible. The last line is a residual contamination from density fluctuations. If the two biases have the same redshift dependence, this term would exactly vanish, allowing us to perfectly isolate lensing magnification. In practice however the two biases evolve differently, and a small density contribution remains.

We compare the \newl estimator with the standard estimator used to measure {lensing magnification}
$
\hes= \hat{C}_\ell^{gg}(\zf,\zb),
$
whose expectation value is
\begin{align}
\label{eq:meanst}
\es \equiv &\langle\hes\rangle=\frac{1}{2}b_g(\zf)(2-5s(\zb))C^{\delta\phi}_\ell(\zf,\zb)\\
&+\frac{1}{2}b_g(\zb)(2-5s(\zf))C^{\phi\delta}_\ell(\zf,\zb)\nonumber\\
& +\frac{1}{4}(2-5s(\zf))(2-5s(\zb))C^{\phi\phi}_\ell(\zf,\zb)\nonumber\\
&+b_g(\zf)b_g(\zb)C_\ell^{\delta\delta}(\zf,\zb)\, .\nonumber
\end{align}
The first and third line correspond to the lensing signal that we want to measure. The third line is due to the fact that both the background and foreground galaxies are lensed by the same structures in front of the foreground galaxies. This contribution is absent in $\hec$ because 21\,cm is not lensed. As before, the second line is negligible. Finally the last line represents the contamination from density fluctuations. The standard way of minimising this contamination consists of choosing $\zb$ and $\zf$ sufficiently far away to make it negligible. 

\vspace{0.2cm}

\emph{Contamination and signal-to-noise ratio.}  Let us now study two questions: does {the \newl} estimator reduce the contamination from density fluctuations? And does {\newl} improve the SNR of {lensing magnification}? {By considering $z_b$ and $z_f$ sufficiently far away, the standard estimator minimises the density contamination in the signal. This contamination reappears however in the variance, where it dominates.} We will see that {the \newl} estimator has the advantage of strongly reducing the density contribution \emph{also} in the variance, consequently increasing the SNR.

We split the signal into a {lensing magnification} contribution, that we want to measure, and the contamination from density: $\ec=\ecl+\ecc$ and $\es=\esl+\esc$. The {lensing magnification} contribution corresponds to the terms in Eqs.~\eqref{eq:meancross} and~\eqref{eq:meanst} that involve the lensing potential $\phi$, while the contamination are the terms proportional to $C_\ell^{\delta \delta}$.
To give a quantitative example of how the contamination is reduced for $\hec$, we use the 
specifications of DES~\cite{DESsurvey} and 
HIRAX~\cite{Newburgh:2016mwi}.
For the cosmological parameters, we use throughout the paper the values from Planck~\cite{Aghanim:2018eyx}.
For redshift pairs separated by $\Delta z=0.25$, we find typically that the contamination {in the signal} is about 1\% for $\hec$, whereas it is 30-40\% for $\hes$. A figure can be found in the Appendix. The {\newl} estimator $\hec$ allows us therefore to extract {lensing magnification} from closer pairs than $\hes$. This is due to a double suppression of the contamination in $\hec$: firstly because the density correlation quickly decreases with redshift separation {(similarly to the standard estimator)}, and secondly by the bias difference. The second suppression is especially effective at small redshift separation, when the bias has not evolved much between $z_f$ and $z_b$. 

{We now calculate the covariance of {the \newl} estimator. We first concentrate on the cosmic variance contribution. The full expression is given in the Appendix and used for the forecasts.} For illustration we discuss here the
diagonal part of the covariance, given by $\zf'=\zf$ and $\zb'=\zb$, corresponding to the variance for the redshift pair $(\zf,\zb)$. It is dominated by the density contribution taken at the same redshift.
We obtain for the standard estimator
\begin{align}
{\rm var}\big[ \hes(\zf,\zb)\big]\simeq\frac{b^2_g(\zf)b^2_g(\zb)}{(2\ell+1)f_{\rm sky}}C_\ell^{\delta\delta}(\zf)C_\ell^{\delta\delta}(\zb)\, ,
\end{align}
and for {\newl} 
\begin{align}
\label{eq:varec}
&{\rm var}\big[ \hec(\zf,\zb)\big]\simeq\frac{1}{(2\ell+1)f_{\rm sky}}\\
&\times\big[b_\im(\zf)b_g(\zb)-b_g(\zf)b_\im(\zb)\big]^2C_\ell^{\delta\delta}(\zf)C_\ell^{\delta\delta}(\zb)\, .\nonumber
\end{align}
This confirms that $\hec$ has the advantage of suppressing the density contribution not only in the mean of the estimator, but also in its variance, thanks to the bias difference which appears in Eq.~\eqref{eq:varec}. 
{The cosmic variance suppression made explicit by \newl is, as usual, also implicitly present in the full likelihood that combines all possible correlations -- indeed, in terms of SNR the likelihood cannot be beaten.}
{However, the \newl estimator is nearly optimal, it is more compact, and it has the additional advantage to reduce the density contamination also in the signal.}

\begin{figure}[t]
\begin{center}
\includegraphics[width=0.5\textwidth]{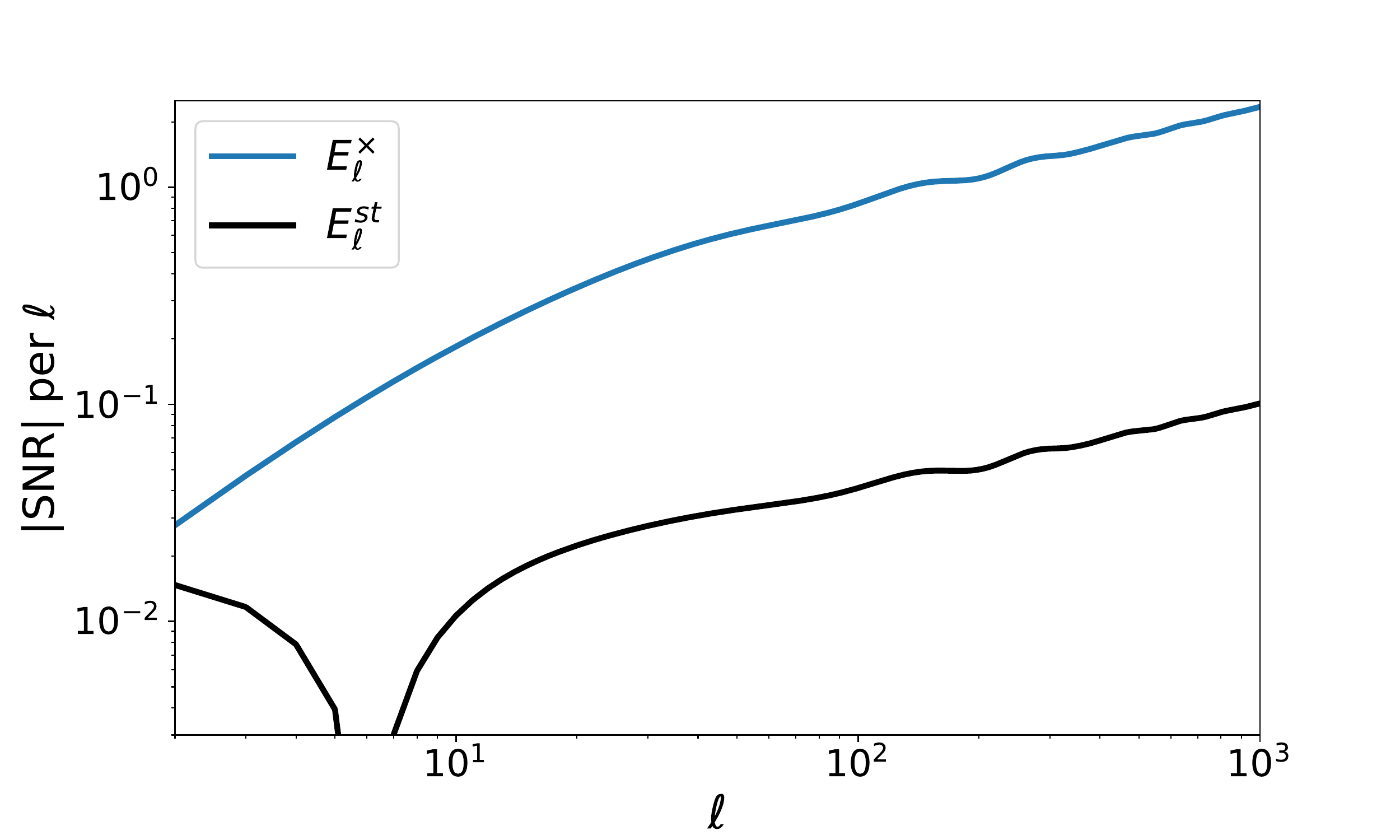}\\
\includegraphics[width=0.45\textwidth]{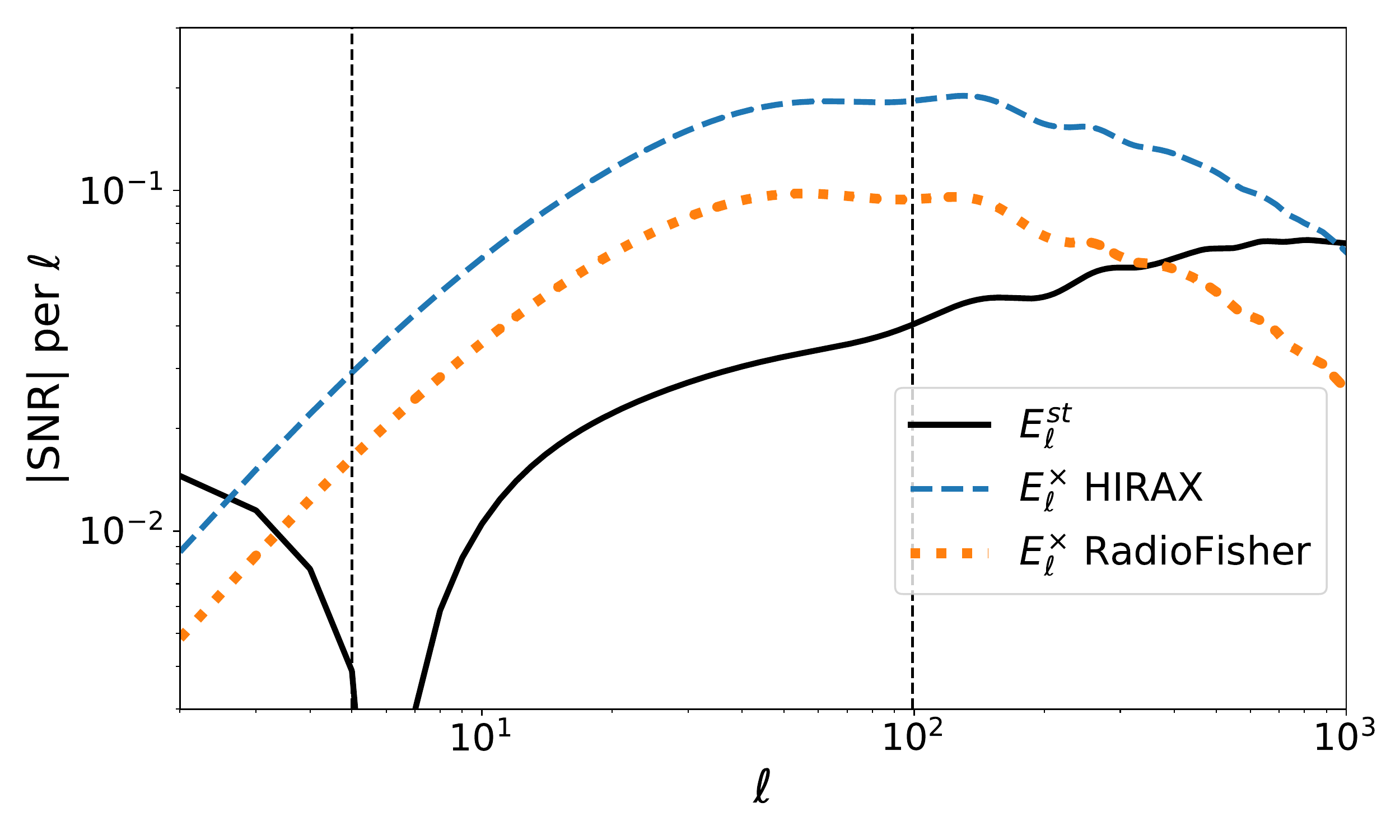}
\end{center}
\caption{\emph{Top panel}: $|{\rm SNR}|$ per $\ell$ mode for {DES ($\es$) and DES$\times$HIRAX ($\ec$)}  with cosmic variance only, for the redshift pair $z_f = 0.8$ and $z_b = 1.3$. \emph{Bottom panel}: Same as the top panel but including shot-noise and thermal noise using the {two} cases discussed in the text.}
\label{fig:SN_ell}
\end{figure}

In Fig.~\ref{fig:SN_ell} (top panel) we plot the SNR per multipole $\ell$ of $\hes$ and $\hec$ for $\zf=0.8$ and $\zb=1.3$, for DES$\times$HIRAX. We assume a sky coverage of 5000 deg$^2$ for both estimators, since HIRAX will overlap with DES. For this case, the contamination is less than 0.01$\%$ in both estimators, so that the signal is simply given by the {lensing magnification}. {Note that the {lensing magnification} in $\hes$ contains both $C_\ell^{\phi\phi}$ and $C_\ell^{\delta\phi}$, which cancel around $\ell=6$. In $\hec$ only $C_\ell^{\delta\phi}$ is present.}
The cumulative SNR, for this redshift pair, from $\ell_{\rm min}=\pi/\theta_{\rm sky} =5 $ to $\ell_{\rm max}=1000$ is $2.4$ for $\hes$ and 54 for $\hec$. If we reduce $\ell_{\rm max}$ to 200, to exclude non-linear scales, the cumulative SNR is $0.6$ for $\hes$ and $12$ for $\hec$.
{The \newl} estimator therefore improves the detection of {lensing magnification} by a factor of $\sim20$ with respect to the conventional method.

The SNR calculated above corresponds to a survey which is cosmic variance limited over the whole range of multipoles. In reality, two additional sources of errors contribute to the variance. First, galaxies are discrete objects, which generate a shot noise contribution to the variance. Shot noise affects both the galaxy number counts and 21\,cm intensity mapping. However, for the latter it has been shown that shot noise is always negligible with respect to the interferometer noise~\cite{Castorina:2016bfm}.  As a consequence we simply replace in the expression for the covariance
$
C_\ell^{gg}(z,z') \rightarrow C_\ell^{gg}(z,z')+\delta_{z,z'}/\bar n(z)\, ,
$
where $\bar n$ denotes the mean number of galaxies per redshift bin and per steradian. 

For interferometer noise, we concentrate on HIRAX which 
will measure the neutral hydrogen distribution 
in the redshift range of $z \sim$ 0.8 to 2.5 covering 15000 square degrees of the southern sky \cite{Newburgh:2016mwi}. 
In the literature we can find several expressions for the noise: two widely used prescriptions are \cite{Bull_HI_noise} and \cite{Zaldarriaga:2003du, Pourtsidou:2013hea}. We discuss them in some detail in the Appendix. We find that for the specific case of HIRAX the two expressions differ by four orders of magnitude, which has a significant impact on the forecasts. The main difference is that~\cite{Bull_HI_noise} assumes that each field of view is observed sequentially, whereas \cite{Zaldarriaga:2003du, Pourtsidou:2013hea} assume that the whole sky is observed at once.
In our forecasts, we show the results for the first scenario {since the second one is very optimistic and not achievable with near future surveys. We label the first scenario as ``RadioFisher''}. In addition, preliminary simulations of the HIRAX interferometer noise based on \cite{Shaw:2013wza,Shaw:2014khi} find a noise curve which is about a factor 10 better than {RadioFisher}. We also include results for this case, that we label as {``HIRAX''}.

In Fig.~\ref{fig:SN_ell} (bottom panel), we plot the SNR for $\hec$ and $\hes$ 
including shot noise and interferometer noise, for the {two cases} discussed above. We use the galaxy number density from DES~\cite{Font-Ribera:2013rwa}. We see that shot noise and interferometer noise significantly reduce the SNR of $\hec$ at large $\ell$. Since cosmic variance is larger for $\hes$, the impact of shot noise is less relevant for this estimator. 
The two vertical lines in Fig.~\ref{fig:SN_ell} correspond respectively to $\ell_{\rm min}=\pi/\theta_{\rm sky}= 5$ and $\ell_{\rm min}=\pi/\theta_{\rm FOV}= 99$. The latter applies in the case where the calibration of each field of view (FOV) is not known, such that only modes smaller than the FOV of the interferometer can be observed.
The cumulative SNR up to $\ell_{\rm max}=1000$ is $3.9$ for $\hec$ and $2.0$ for $\hes$. For $\ell_{\rm max}=200$ we find $2.4$ for $\hec$ and $0.6$ for $\hes$. This improvement by a factor of 4 allows a marginal detection of lensing with $\hec$ from a \emph{single} redshift pair, for which $\hes$ cannot detect anything. This is particularly useful to follow the redshift evolution of dark energy or modified gravity.

\vspace{0.2cm}

\emph{Forecasts on the lensing amplitude $A_L$.} As an application of {\newl}, we forecast the precision with which we will be able to measure the amplitude of the lensing potential $\phi$. For this we replace $\phi\rightarrow A_L\cdot\phi$ in Eq.~\eqref{eq:phi} and we forecast the error on $A_L$, with fiducial value $A_L=1$. 
We fix all cosmological and astrophysical parameters to their fiducial value, and we compute the Fisher element for $A_L$
\begin{align}
\label{eq:Fisher}
\FF^{\times}_{A_L}=&\sum_{\zf,\zb,\zf'\zb'}\sum_{\ell=\ell_{\rm min}}^{\ell_{\rm max}}\frac{\partial \ec}{\partial A_L}(\zf,\zb)\\
&\times{\rm Cov}^{-1}\big[\hec(\zf,\zb)\hec(\zf',\zb') \big]
\frac{\partial \ec}{\partial A_L}(\zf',\zb')\, ,\nonumber
\end{align}
and similarly for $\es$. We use Gaussian redshift bins of width $\sigma_z=0.05$, spaced by $\Delta z=0.1$, and we sum over all possible pairs of redshifts accessible in the survey, with the condition that the contamination for each pair is below 1\%. This means that in both cases, we exclude pairs with redshift difference smaller than 0.3. For $\hec$ we could consider pairs down to a difference of $0.25$ but that would require finer redshift bins than the current analysis.
This negligible amount of contamination ensures that we measure $A_L$ in a model-independent way, i.e.\ without having to model the evolution of density fluctuations.

\begin{figure}[t]
\begin{center}
\includegraphics[width=0.5\textwidth]{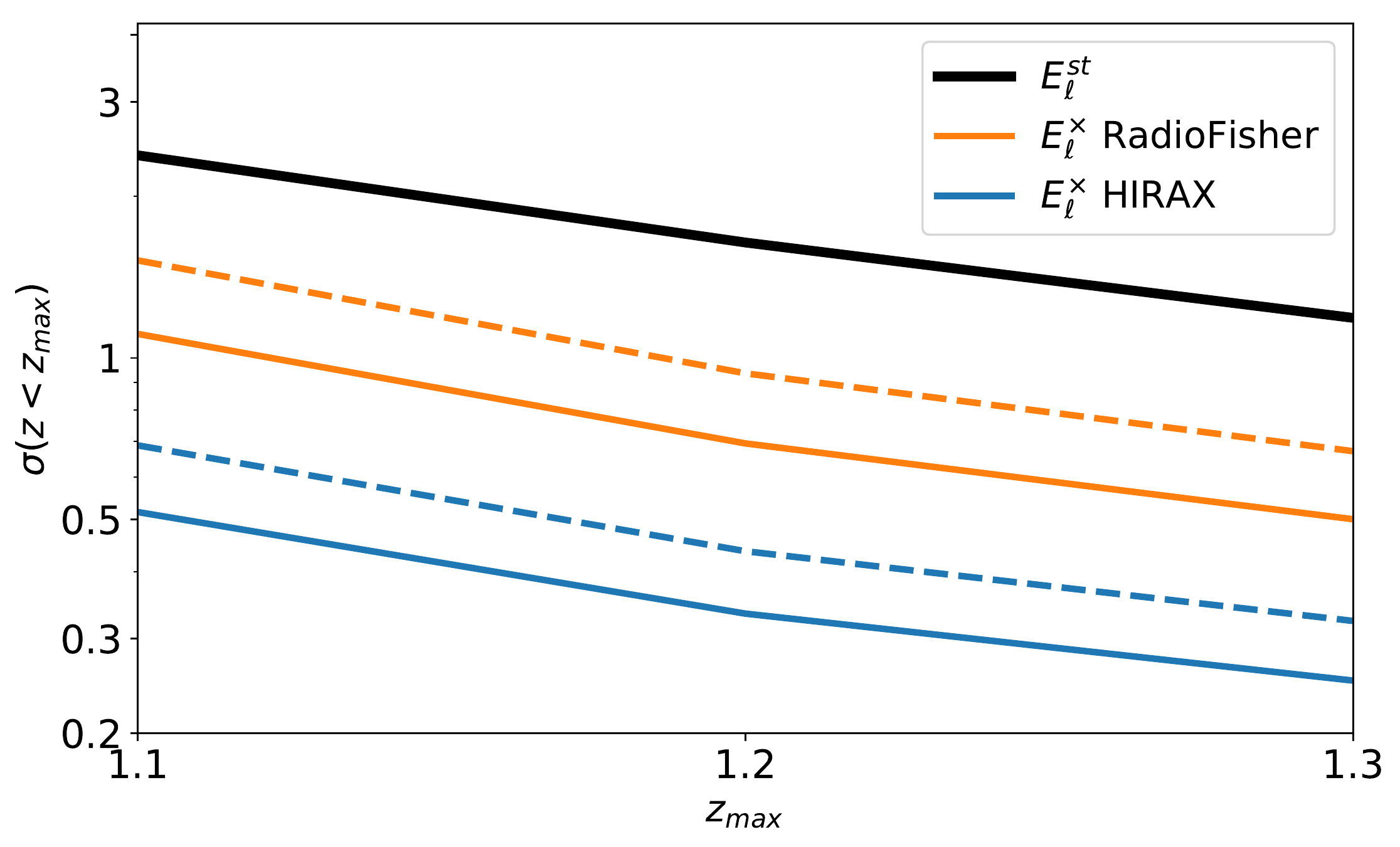}\\
\includegraphics[width=0.5\textwidth]{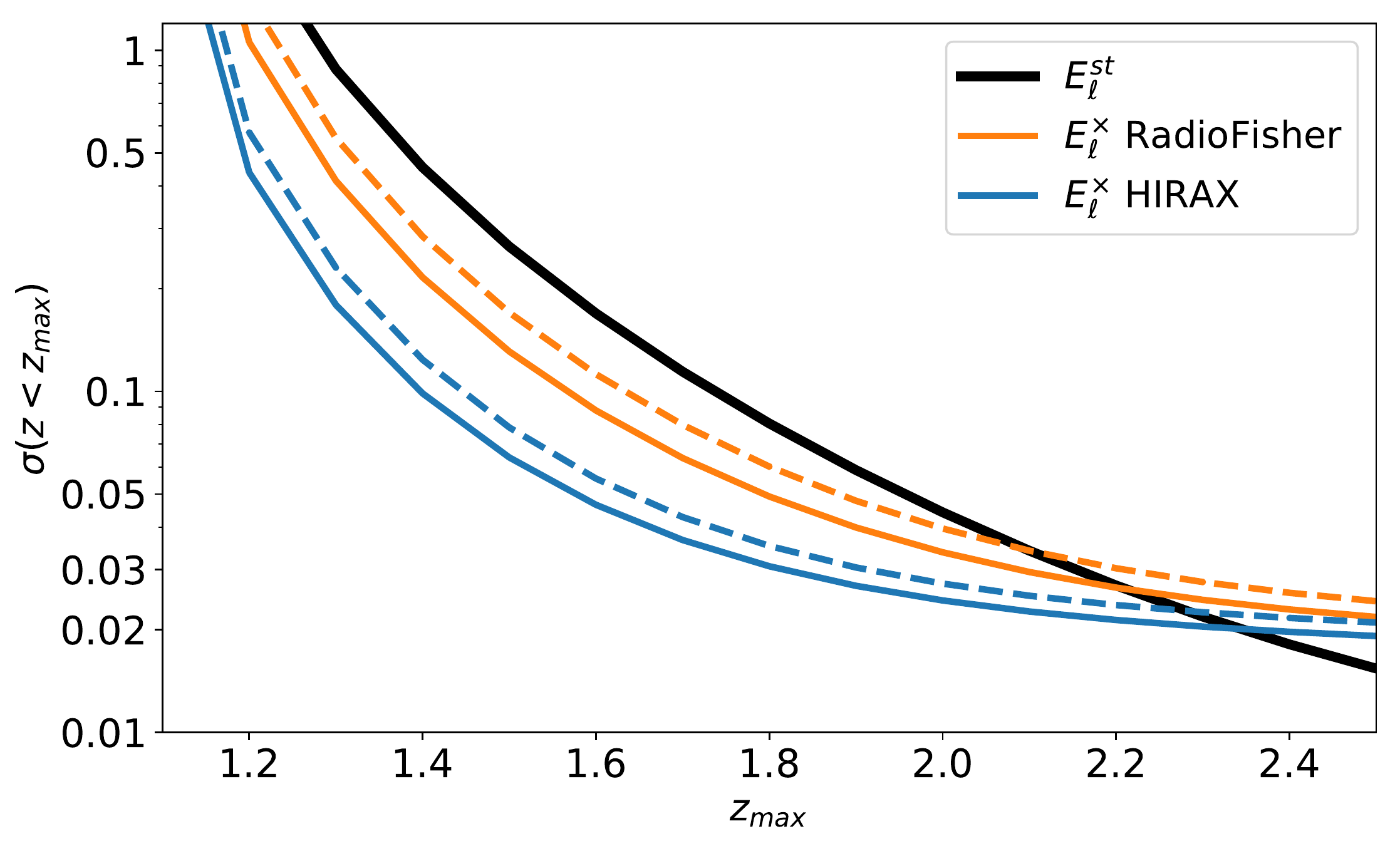}
\end{center}
\caption{Forecasted uncertainty on the lensing amplitude $A_L$ for DES$\times$HIRAX (top panel) and Euclid$\times$HIRAX (bottom panel) as a function of $z_{\rm max}$ for the {two} cases of thermal noise discussed in the text. Solid lines correspond to $\ell_{\rm min}=5$, and dashed lines to $\ell_{\rm min}=\pi/\theta_{\rm FOV}\sim 50-100$ depending on redshift.}
\label{fig:sigma}
\end{figure}

We compute the Fisher matrix for both estimators for two combinations of surveys: DES$\times$HIRAX, and Euclid (photometric)$\times$HIRAX. For the second case, we use $f_{\rm sky}=2/3 \, f_{\rm sky\, Euclid}$ (where $f_{\rm sky\, Euclid} \simeq 0.36$) for $\hec$ since HIRAX will not completely overlap with Euclid. We focus on the common redshift range, which is $z\in[0.8,1.3]$ for DES$\times$HIRAX and $z\in[0.8,2.5]$ for Euclid$\times$HIRAX. For the standard estimator we also use this range as a point of comparison. Our goal is to understand how $\hec$ improves over $\hes$ over the common range. An ideal analysis would then combine $\hec$ over the common range, with $\hes$ over the rest of the optical range. We restrict the $\ell$-range to $\ell_{\rm max}=200$, {which is comparable to \cite{Ho2012,Camacho2019} but more conservative to ensure that linear theory is valid.}

In Fig.~\ref{fig:sigma} we show the precision on the measurement of $A_L$, $\sigma_{A_L}=1/\sqrt{\mathcal{F}_{A_L}}$, as a function of the maximum redshift included in Eq.~\eqref{eq:Fisher}, for DES$\times$HIRAX (top panel) and Euclid$\times$HIRAX (bottom panel). For the case $\ell_{\rm min}=5$, we see that our estimator allows to measure the lensing amplitude with a precision of 0.3 (using the HIRAX noise curve). This corresponds to an improvement of a factor 4.7 compared to the standard estimator in that same redshift range. If we use $\ell_{\rm min}=\pi/\theta_{\rm FOV}$, the improvement is slightly smaller, but still interesting: 3.7 in the realistic case. This clearly shows that, over the common redshift range, our estimator is an excellent tool to measure gravitational lensing. To reach a similar improvement using a single survey we would need to increase the sky coverage by a factor 14 in that redshift range.

Using Euclid$\times$HIRAX, we reach $\sigma_{A_L}=0.05$ for $z_{\rm max}=1.6$ and $\sigma_{A_L}=0.02$ if we include pairs up to $z_{\rm max}=2.5$ (realistic case, and $\ell_{\rm min}=5$). Comparing with the standard estimator, in the same redshift range, we find an improvement of a factor 3.6 at $z_{\rm max}=1.6$, whereas at $z_{\rm max}=2.5$ the standard estimator is slightly better. This is due to the fact that at high redshift, the signal becomes larger for $\hes$ than for $\hec$, due to the lensing-lensing contribution, $C_\ell^{\phi\phi}$, which strongly increases with redshift and which is present in $\hes$ but not in $\hec$. {The \newl} estimator is therefore mainly valuable at intermediate redshift, where it allows a clean and better measurement of $C_\ell^{\delta\phi}$ on its own. Moreover, combining the two estimators would provide separate measurements of $C_\ell^{\delta\phi}$ and $C_\ell^{\phi\phi}$. This is particularly useful to measure the evolution of the lensing potential, and study the impact of dark energy and modified gravity as a function of redshift.

\emph{Conclusion.} We have constructed a new estimator, {\newl}, to measure {lensing magnification}, using the correlation of galaxy clustering and 21\,cm intensity mapping. {The \newl} estimator improves over standard {lensing magnification} estimators from galaxy clustering in several ways. First, it allows us to significantly reduce the contamination from density correlations, when the foreground and background redshifts are close. 
Second, the cosmic variance of {the \newl} estimator is significantly reduced with respect to the standard estimator. {\newl} improves the SNR by a factor of 30 for cosmic variance limited surveys, relative to the standard estimator. Shot noise and interferometer noise reduce this improvement, but it still reaches a factor 4-5 at intermediate redshift. Finally, as it involves only cross-correlations, {the \newl} estimator {will be robust to most systematics from galaxy and IM surveys, except for correlated foregrounds e.g.\ from the Milky Way which should be projected out by approaches such as \cite{Crocce2016,Elvin-Poole2018}}.
It is also possible to extend the estimator to remove the density contribution completely, a development that we will discuss in a future publication. This {\em letter} already shows that the fundamental idea of combining intensity mapping and galaxy surveys to isolate {lensing magnification} holds considerable promise for upcoming surveys.

\emph{Acknowledgements}
We thank Alkistis Pourtsidou, Stefano Camera and Phil Bull for enlightening us about the intensity mapping noise, Moumita Aich for providing us intensity mapping bias, Devin Crichton and Azadeh Moradinezhad for useful discussions about HIRAX and intensity mapping forecasts. CB, MJ, MK and FL acknowledge funding by the Swiss National Science Foundation. EM acknowledges financial support from the Swiss National Science Foundation through the Marie Heim-V\"ogtlin Fellowship No.~PMPDP2\_171332. KM acknowledges support from the National Research Foundation of South Africa (grant number 98957). WN acknowledges the financial assistance of the South African Radio Astronomy Observatory (SARAO) towards this research (www.ska.ac.za).

\begin{appendix}

\section{Covariances\label{app:covar}}

Here we give the full covariances from cosmic variance for the two estimators. For the standard estimator, $\es$, it reads
\begin{align}
\label{eq:coves}
&{\rm cov}\big[\hes(\zf,\zb)\hes(\zf',\zb') \big]=\frac{1}{(2\ell+1)f_{\rm sky}}\\
&\times \Big[C_\ell^{gg}(\zf,\zf')C_\ell^{gg}(\zb,\zb')+C_\ell^{gg}(\zf,\zb')C_\ell^{gg}(\zf',\zb)\Big]\, .\nonumber
\end{align}
For {the \newl} estimator, $\ec$, we obtain
\begin{align}
\label{eq:covec}
&{\rm cov}\big[\hec(\zf,\zb)\hec(\zf',\zb') \big]=\frac{1}{(2\ell+1)f_{\rm sky}}\\
&\times \Big[C_\ell^{\im\im}(\zf,\zf')C_\ell^{g g}(\zb,\zb')+C_\ell^{gg}(\zf,\zf')C_\ell^{\im\im}(\zb,\zb')\nonumber\\
&-C_\ell^{\im g}(\zf,\zf')C_\ell^{g\im}(\zb,\zb')-C_\ell^{g\im}(\zf,\zf')C_\ell^{\im g}(\zb,\zb')\nonumber\\
&+C_\ell^{g \im}(\zb,\zf')C_\ell^{\im g}(\zf,\zb')+C_\ell^{\im g}(\zb,\zf')C_\ell^{g \im}(\zf,\zb')\nonumber\\
&-C_\ell^{g g}(\zb,\zf')C_\ell^{\im \im}(\zf,\zb')-C_\ell^{\im \im}(\zb,\zf')C_\ell^{gg}(\zf,\zb')\Big]\, .\nonumber
\end{align}

\section{Contamination\label{app:contamination}}

\begin{figure}[!h]
\begin{center}
\includegraphics[width=0.45\textwidth]{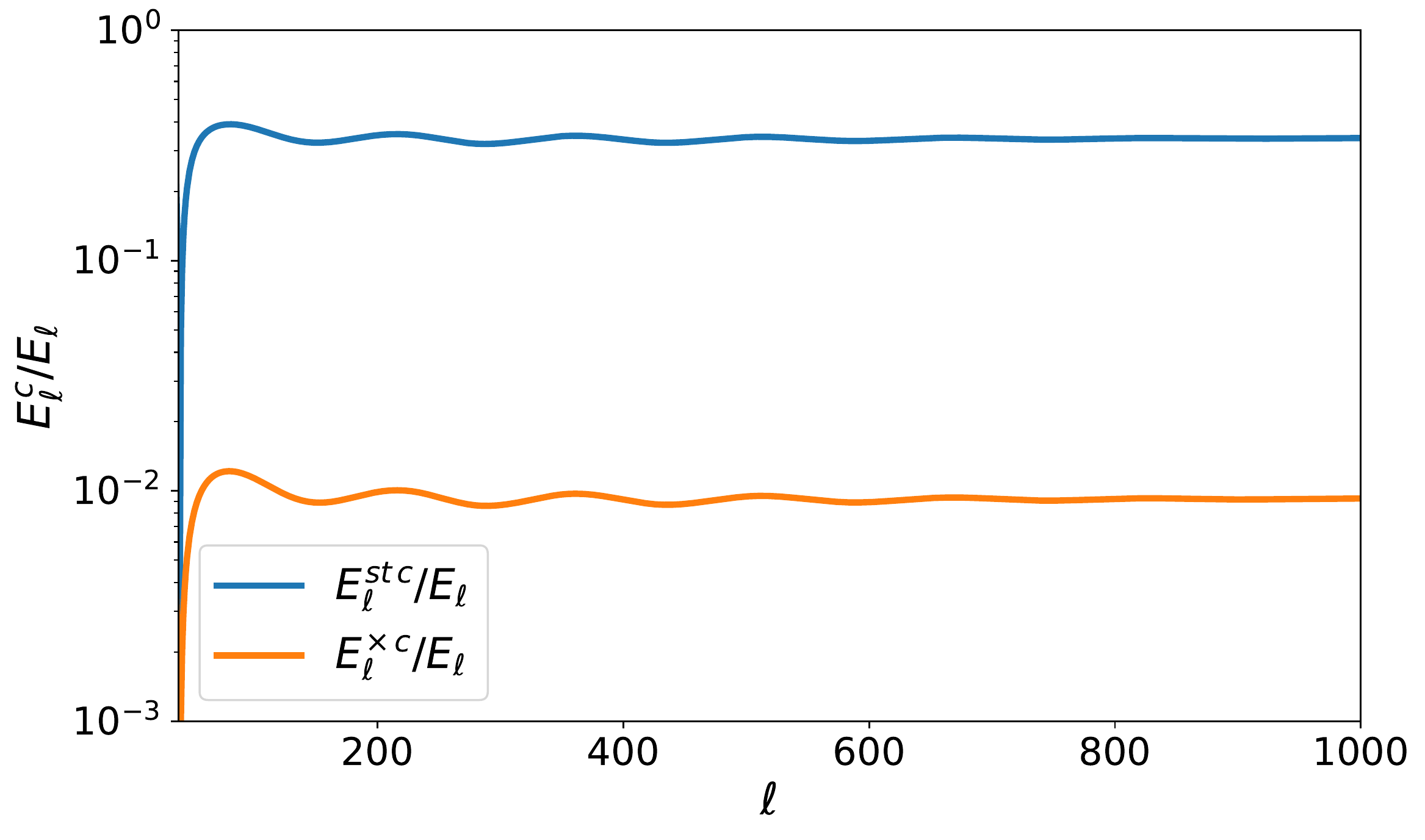}
\end{center}
\caption{Ratio of contamination (the contribution from density-density correlations) to the full estimator for $\ec$ and $\es$ for DES$\times$HIRAX and for $z_f=1$ and $z_b=1.25$.}
\label{fig:contamination}
\end{figure}

In Fig.~\ref{fig:contamination}, we show a plot of the ratio of contamination for the case DES$\times$HIRAX and for the specific redshift pair $z_f=1$ and $z_b=1.25$. We see that GIMCO is much less contaminated by the density contribution than the standard estimator. Note that a similarly high contamination from density would be present in each term of the full likelihood that combines all possible correlations.

\section{Interferometer noise\label{app:noise}}

Reference~\cite{Bull_HI_noise} gives the noise spectrum as
\be
C^{\rm interf}_\ell
(z) = \frac{T_{\rm sys}^2 \, S_{\rm area} \, \lambda(z)^4}{n_{\rm pol} \, t_{\rm tot} \, \Delta \nu \, N_{\rm beam} \, A_{\rm eff}^2 \, \theta _b^2 \, n(u=\ell/2\pi)}
\,,
\label{eq:noiseBULL}
\ee
where $T_{\rm sys}=T_{\rm antenna}+T_{\rm sky}$ is the addition of antenna and sky temperature, $S_{\rm area}= 4\pi f_{\rm sky}$ is the observed area of the sky, $\lambda(z) $ is the observed wavelength of 21cm line at redshift $z$, $n_{\rm pol}$ is the number of polarizations, $t_{\rm tot}$ is the total observation time, $\Delta \nu$ is the frequency bin corresponding to the redshift bin width, $N_{\rm beam}$ is the beam number, $A_{\rm eff}= 0.7 \pi D_{\rm dish}^2/4$ is the effective area of each dish and the factor $0.7$ is the efficiency of the dish, $\theta_b = \lambda / D_{\rm dish}$ is the beam of the telescope, and $n(u)$ is the number density of baselines in the $uv$ plane. Expression~\eqref{eq:noiseBULL} assumes that each field of view (FOV) of the interferometer is observed sequentially. If on the other hand one assumes that the whole sky area is observed at once (instantaneous FOV), we obtain the expression presented in~\cite{Zaldarriaga:2003du, Pourtsidou:2013hea}
\be
C^{\rm interf}_\ell= \frac{(2\pi)^3 \, T_{\rm sys}^2}{\Delta \nu \, t_{\rm tot} \, f_{\rm cover}^2 \, \ell_{\rm max}(\nu)^2} \,,
\label{eq:noiseAL}
\ee 
where $\ell_{\rm max} = 2\pi D_{tel}/\lambda(z)$ and $D_{\rm tel }$ is the diameter of the telescope array, $f_{\rm cover} = N_{\rm dish} A_{\rm eff} / (\pi D^2_{\rm tel}/4)$ is the effective collecting area of the telescope. For HIRAX the difference between the two noise curves is of the order of $10^4$. However, we do not show the results for this noise as it appears too optimistic for near future surveys.
\section{Galaxy and IM specifications\label{app:bias}}
\begin{figure}[!h]
    \centering
    \includegraphics[width=0.45\textwidth]{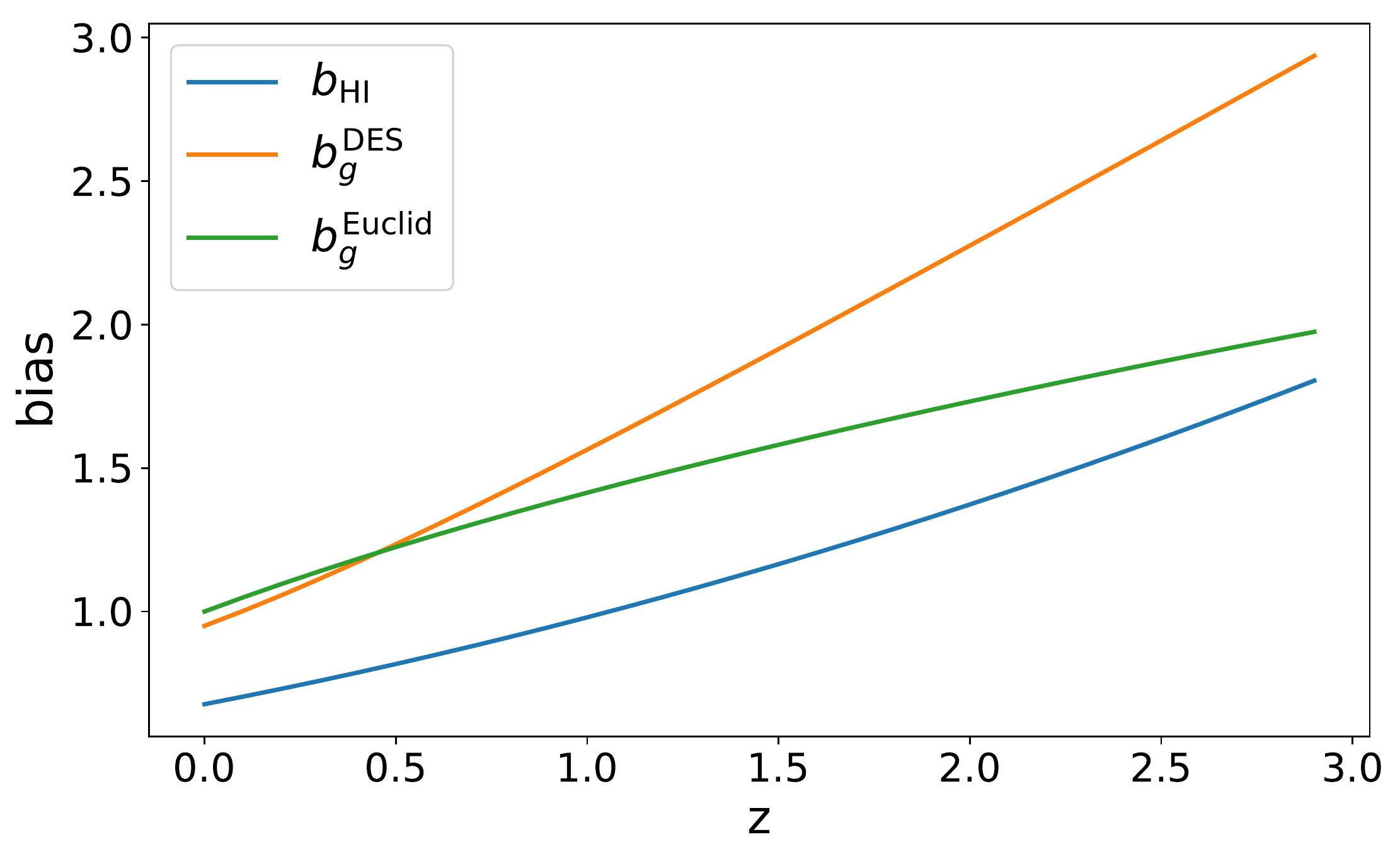}
    \caption{HIRAX, DES, and Euclid bias as a function of redshift.}
    \label{fig:bias}
\end{figure}

\begin{figure}[!h]
    \centering
    \includegraphics[width=0.45\textwidth]{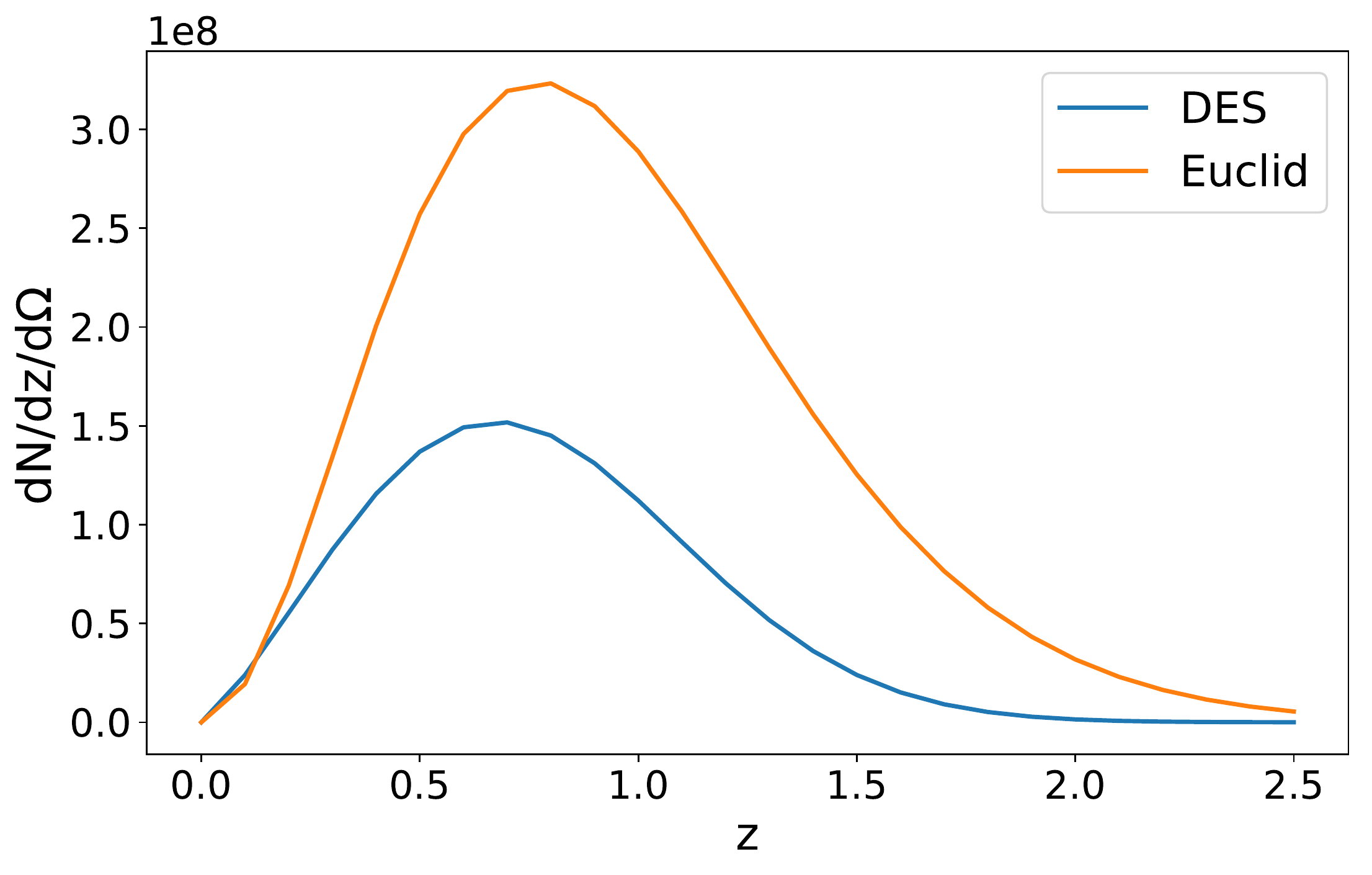}
    \caption{Number density of galaxies per steradian$^2$ per redshift for DES and Euclid surveys.}
    \label{fig:numberdensity}
\end{figure}

Here we give a list of the specifications used for galaxy and IM surveys.
\\
\textbf{DES}
\\
For DES survey, we use the following relation for bias~\cite{Font-Ribera:2013rwa}
\be b_g^{\rm DES}(z) = 0.95 \, \frac{D(z=0)}{D(z)} \, ,\ee
where $D$ is the growth factor. For number density, we use the following relation ~\cite{Font-Ribera:2013rwa}
\be \frac{dN}{dz d\Omega} = N \left(z/z_*\right)^{\alpha} \exp[\left(-z/z_*\right)^{\beta}]\, ,\ee
where $\alpha=1.25$, $\beta=2.29$, $z_{*}=0.88$, and $N$ is the normalization factor given by
\be N= \frac{n_{tot}}{ (z_*/\beta) \Gamma\left((1+\alpha)/\beta\right) }\, ,\ee
where $\Gamma$ is the Gamma function and $n_{tot}$ is the total number of galaxies per arcmin$^2$. For DES, $n_{tot}=12 \, [\rm arcmin ^{-2}]$.
We use a constant value, $s=0.52$ for the magnification bias following \cite{Garcia-Fernandez:2016oud}.
\\
\textbf{Euclid}
\\
For Euclid photometric survey, we use the following specifications \cite{Montanari:2015rga}
\begin{eqnarray}
&&b_g^{\rm Euclid}(z) = b_0 \sqrt{1+z}\\ \nonumber
&&s = s_0 + s_1 z + s_2 z^2 + s_3 z^3\\ \nonumber
&&  \frac{dN}{dz d\Omega} = 2\times 10^9 z^2 \exp \left[(-(z/z_0)^{1.5})\right]\, ,
\end{eqnarray}
where $b_0=1$, $s_0 = 0.1194$, $s_1 = 0.2122$, $s_2 = - 0.0671$, and $s_3 = 0.1031$.
The normalization factor in number density is determined by demanding a total number of $n_{tot}= 30$ arcmin$^{-2}$ galaxies from $z=0$ to $z=2.5$. We find a different normalization factor from \cite{Montanari:2015rga}.
\\
\textbf{HIRAX}
\\
For HIRAX bias, we use the fitting formula from RadioFisher code\cite{radiofisher}
\be 
b_{\rm HI}(z) = 0.68(1 + 0.38 z + 0.067 z^2)
\ee
In Fig.~\ref{fig:bias}, we show bias as a function of redshift for HIRAX, Euclid and DES surveys and in Fig.~\ref{fig:numberdensity}, we show $\rm dN/dz/d\Omega$ for Euclid and DES.
%
\section{Redshift pairs}
Here we give explicitly the list of redshift pairs that has been used for Fisher analysis. For both $\es$ and $\ec$, we use redshift pairs in the common redshift range of the galaxy and intensity mapping survey that are far enough so that the density contamination is negligible which translates to $(z_b - z_f) \ge 0.3$ . The center of redshift bins are separated by $0.1$. For DES$\times$HIRAX, we use the following pairs
\begin{align} \{(z_f,z_b)\}=& \{(0.8,1.1),(0.8,1.2),(0.81.3),(0.9,1.2)\\ \nonumber
& ,(0.9,1.3),(1,1.3)\} \, , \end{align}
and for Euclid $\times$ HIRAX, we use
\begin{align} \{(z_f,z_b)\}=& \{(0.8,1.1),(0.8,1.2),...,(0.8,2.5)\\ \nonumber
&,(0.9,1.2),...,(0.9,2.5)
,...,(2.2,2.5)\}. \end{align}
\end{appendix}

\bibliography{21crossgalbib}

\begin{thebibliography}{41}%
\makeatletter
\providecommand \@ifxundefined [1]{%
 \@ifx{#1\undefined}
}%
\providecommand \@ifnum [1]{%
 \ifnum #1\expandafter \@firstoftwo
 \else \expandafter \@secondoftwo
 \fi
}%
\providecommand \@ifx [1]{%
 \ifx #1\expandafter \@firstoftwo
 \else \expandafter \@secondoftwo
 \fi
}%
\providecommand \natexlab [1]{#1}%
\providecommand \enquote  [1]{``#1''}%
\providecommand \bibnamefont  [1]{#1}%
\providecommand \bibfnamefont [1]{#1}%
\providecommand \citenamefont [1]{#1}%
\providecommand \href@noop [0]{\@secondoftwo}%
\providecommand \href [0]{\begingroup \@sanitize@url \@href}%
\providecommand \@href[1]{\@@startlink{#1}\@@href}%
\providecommand \@@href[1]{\endgroup#1\@@endlink}%
\providecommand \@sanitize@url [0]{\catcode `\\12\catcode `\$12\catcode
  `\&12\catcode `\#12\catcode `\^12\catcode `\_12\catcode `\%12\relax}%
\providecommand \@@startlink[1]{}%
\providecommand \@@endlink[0]{}%
\providecommand \url  [0]{\begingroup\@sanitize@url \@url }%
\providecommand \@url [1]{\endgroup\@href {#1}{\urlprefix }}%
\providecommand \urlprefix  [0]{URL }%
\providecommand \Eprint [0]{\href }%
\providecommand \doibase [0]{http://dx.doi.org/}%
\providecommand \selectlanguage [0]{\@gobble}%
\providecommand \bibinfo  [0]{\@secondoftwo}%
\providecommand \bibfield  [0]{\@secondoftwo}%
\providecommand \translation [1]{[#1]}%
\providecommand \BibitemOpen [0]{}%
\providecommand \bibitemStop [0]{}%
\providecommand \bibitemNoStop [0]{.\EOS\space}%
\providecommand \EOS [0]{\spacefactor3000\relax}%
\providecommand \BibitemShut  [1]{\csname bibitem#1\endcsname}%
\let\auto@bib@innerbib\@empty
\bibitem [{\citenamefont {Bacon}\ \emph {et~al.}(2000)\citenamefont {Bacon},
  \citenamefont {Refregier},\ and\ \citenamefont {Ellis}}]{Bacon:2000sy}%
  \BibitemOpen
  \bibfield  {author} {\bibinfo {author} {\bibfnamefont {D.~J.}\ \bibnamefont
  {Bacon}}, \bibinfo {author} {\bibfnamefont {A.~R.}\ \bibnamefont
  {Refregier}}, \ and\ \bibinfo {author} {\bibfnamefont {R.~S.}\ \bibnamefont
  {Ellis}},\ }\href {\doibase 10.1046/j.1365-8711.2000.03851.x} {\bibfield
  {journal} {\bibinfo  {journal} {Mon. Not. Roy. Astron. Soc.}\ }\textbf
  {\bibinfo {volume} {318}},\ \bibinfo {pages} {625} (\bibinfo {year}
  {2000})},\ \Eprint {http://arxiv.org/abs/astro-ph/0003008}
  {arXiv:astro-ph/0003008 [astro-ph]} \BibitemShut {NoStop}%
\bibitem [{\citenamefont {Wittman}\ \emph {et~al.}(2000)\citenamefont
  {Wittman}, \citenamefont {Tyson}, \citenamefont {Kirkman}, \citenamefont
  {Dell'Antonio},\ and\ \citenamefont {Bernstein}}]{Wittman:2000tc}%
  \BibitemOpen
  \bibfield  {author} {\bibinfo {author} {\bibfnamefont {D.~M.}\ \bibnamefont
  {Wittman}}, \bibinfo {author} {\bibfnamefont {J.~A.}\ \bibnamefont {Tyson}},
  \bibinfo {author} {\bibfnamefont {D.}~\bibnamefont {Kirkman}}, \bibinfo
  {author} {\bibfnamefont {I.}~\bibnamefont {Dell'Antonio}}, \ and\ \bibinfo
  {author} {\bibfnamefont {G.}~\bibnamefont {Bernstein}},\ }\href {\doibase
  10.1038/35012001} {\bibfield  {journal} {\bibinfo  {journal} {Nature}\
  }\textbf {\bibinfo {volume} {405}},\ \bibinfo {pages} {143} (\bibinfo {year}
  {2000})},\ \Eprint {http://arxiv.org/abs/astro-ph/0003014}
  {arXiv:astro-ph/0003014 [astro-ph]} \BibitemShut {NoStop}%
\bibitem [{\citenamefont {van Waerbeke}\ \emph {et~al.}(2000)\citenamefont {van
  Waerbeke} \emph {et~al.}}]{vanWaerbeke:2000rm}%
  \BibitemOpen
  \bibfield  {author} {\bibinfo {author} {\bibfnamefont {L.}~\bibnamefont {van
  Waerbeke}} \emph {et~al.},\ }\href@noop {} {\bibfield  {journal} {\bibinfo
  {journal} {Astron. Astrophys.}\ }\textbf {\bibinfo {volume} {358}},\ \bibinfo
  {pages} {30} (\bibinfo {year} {2000})},\ \Eprint
  {http://arxiv.org/abs/astro-ph/0002500} {arXiv:astro-ph/0002500 [astro-ph]}
  \BibitemShut {NoStop}%
\bibitem [{\citenamefont {Kaiser}\ \emph {et~al.}(2000)\citenamefont {Kaiser},
  \citenamefont {Wilson},\ and\ \citenamefont {Luppino}}]{Kaiser:2000if}%
  \BibitemOpen
  \bibfield  {author} {\bibinfo {author} {\bibfnamefont {N.}~\bibnamefont
  {Kaiser}}, \bibinfo {author} {\bibfnamefont {G.}~\bibnamefont {Wilson}}, \
  and\ \bibinfo {author} {\bibfnamefont {G.~A.}\ \bibnamefont {Luppino}},\
  }\href@noop {} {\  (\bibinfo {year} {2000})},\ \Eprint
  {http://arxiv.org/abs/astro-ph/0003338} {arXiv:astro-ph/0003338 [astro-ph]}
  \BibitemShut {NoStop}%
\bibitem [{\citenamefont {Troxel}\ \emph {et~al.}(2018)\citenamefont {Troxel}
  \emph {et~al.}}]{Troxel:2017xyo}%
  \BibitemOpen
  \bibfield  {author} {\bibinfo {author} {\bibfnamefont {M.~A.}\ \bibnamefont
  {Troxel}} \emph {et~al.} (\bibinfo {collaboration} {DES}),\ }\href {\doibase
  10.1103/PhysRevD.98.043528} {\bibfield  {journal} {\bibinfo  {journal} {Phys.
  Rev.}\ }\textbf {\bibinfo {volume} {D98}},\ \bibinfo {pages} {043528}
  (\bibinfo {year} {2018})},\ \Eprint {http://arxiv.org/abs/1708.01538}
  {arXiv:1708.01538 [astro-ph.CO]} \BibitemShut {NoStop}%
\bibitem [{\citenamefont {{Hildebrandt}}\ \emph {et~al.}(2018)\citenamefont
  {{Hildebrandt}}, \citenamefont {{K{\"o}hlinger}}, \citenamefont {{van den
  Busch}}, \citenamefont {{Joachimi}}, \citenamefont {{Heymans}}, \citenamefont
  {{Kannawadi}}, \citenamefont {{Wright}}, \citenamefont {{Asgari}},
  \citenamefont {{Blake}},\ and\ \citenamefont
  {{Hoekstra}}}]{Hildebrandt:2018yau}%
  \BibitemOpen
  \bibfield  {author} {\bibinfo {author} {\bibfnamefont {H.}~\bibnamefont
  {{Hildebrandt}}}, \bibinfo {author} {\bibfnamefont {F.}~\bibnamefont
  {{K{\"o}hlinger}}}, \bibinfo {author} {\bibfnamefont {J.~L.}\ \bibnamefont
  {{van den Busch}}}, \bibinfo {author} {\bibfnamefont {B.}~\bibnamefont
  {{Joachimi}}}, \bibinfo {author} {\bibfnamefont {C.}~\bibnamefont
  {{Heymans}}}, \bibinfo {author} {\bibfnamefont {A.}~\bibnamefont
  {{Kannawadi}}}, \bibinfo {author} {\bibfnamefont {A.~H.}\ \bibnamefont
  {{Wright}}}, \bibinfo {author} {\bibfnamefont {M.}~\bibnamefont {{Asgari}}},
  \bibinfo {author} {\bibfnamefont {C.}~\bibnamefont {{Blake}}}, \ and\
  \bibinfo {author} {\bibfnamefont {H.}~\bibnamefont {{Hoekstra}}},\
  }\href@noop {} {\bibfield  {journal} {\bibinfo  {journal} {arXiv e-prints}\
  ,\ \bibinfo {eid} {arXiv:1812.06076}} (\bibinfo {year} {2018})},\ \Eprint
  {http://arxiv.org/abs/1812.06076} {arXiv:1812.06076 [astro-ph.CO]}
  \BibitemShut {NoStop}%
\bibitem [{\citenamefont {Hamana}\ \emph {et~al.}(2019)\citenamefont {Hamana}
  \emph {et~al.}}]{Hamana:2019etx}%
  \BibitemOpen
  \bibfield  {author} {\bibinfo {author} {\bibfnamefont {T.}~\bibnamefont
  {Hamana}} \emph {et~al.},\ }\href@noop {} {\  (\bibinfo {year} {2019})},\
  \Eprint {http://arxiv.org/abs/1906.06041} {arXiv:1906.06041 [astro-ph.CO]}
  \BibitemShut {NoStop}%
\bibitem [{\citenamefont {Jee}\ \emph {et~al.}(2016)\citenamefont {Jee},
  \citenamefont {Tyson}, \citenamefont {Hilbert}, \citenamefont {Schneider},
  \citenamefont {Schmidt},\ and\ \citenamefont {Wittman}}]{Jee:2015jta}%
  \BibitemOpen
  \bibfield  {author} {\bibinfo {author} {\bibfnamefont {M.~J.}\ \bibnamefont
  {Jee}}, \bibinfo {author} {\bibfnamefont {J.~A.}\ \bibnamefont {Tyson}},
  \bibinfo {author} {\bibfnamefont {S.}~\bibnamefont {Hilbert}}, \bibinfo
  {author} {\bibfnamefont {M.~D.}\ \bibnamefont {Schneider}}, \bibinfo {author}
  {\bibfnamefont {S.}~\bibnamefont {Schmidt}}, \ and\ \bibinfo {author}
  {\bibfnamefont {D.}~\bibnamefont {Wittman}},\ }\href {\doibase
  10.3847/0004-637X/824/2/77} {\bibfield  {journal} {\bibinfo  {journal}
  {Astrophys. J.}\ }\textbf {\bibinfo {volume} {824}},\ \bibinfo {pages} {77}
  (\bibinfo {year} {2016})},\ \Eprint {http://arxiv.org/abs/1510.03962}
  {arXiv:1510.03962 [astro-ph.CO]} \BibitemShut {NoStop}%
\bibitem [{\citenamefont {Scranton}\ \emph {et~al.}(2005)\citenamefont
  {Scranton} \emph {et~al.}}]{Scranton:2005ci}%
  \BibitemOpen
  \bibfield  {author} {\bibinfo {author} {\bibfnamefont {R.}~\bibnamefont
  {Scranton}} \emph {et~al.} (\bibinfo {collaboration} {SDSS}),\ }\href
  {\doibase 10.1086/431358} {\bibfield  {journal} {\bibinfo  {journal}
  {Astrophys. J.}\ }\textbf {\bibinfo {volume} {633}},\ \bibinfo {pages} {589}
  (\bibinfo {year} {2005})},\ \Eprint {http://arxiv.org/abs/astro-ph/0504510}
  {arXiv:astro-ph/0504510 [astro-ph]} \BibitemShut {NoStop}%
\bibitem [{\citenamefont {Duncan}\ \emph {et~al.}(2014)\citenamefont {Duncan},
  \citenamefont {Joachimi}, \citenamefont {Heavens}, \citenamefont {Heymans},\
  and\ \citenamefont {Hildebrandt}}]{Duncan:2013haa}%
  \BibitemOpen
  \bibfield  {author} {\bibinfo {author} {\bibfnamefont {C.}~\bibnamefont
  {Duncan}}, \bibinfo {author} {\bibfnamefont {B.}~\bibnamefont {Joachimi}},
  \bibinfo {author} {\bibfnamefont {A.}~\bibnamefont {Heavens}}, \bibinfo
  {author} {\bibfnamefont {C.}~\bibnamefont {Heymans}}, \ and\ \bibinfo
  {author} {\bibfnamefont {H.}~\bibnamefont {Hildebrandt}},\ }\href {\doibase
  10.1093/mnras/stt2060} {\bibfield  {journal} {\bibinfo  {journal} {Mon. Not.
  Roy. Astron. Soc.}\ }\textbf {\bibinfo {volume} {437}},\ \bibinfo {pages}
  {2471} (\bibinfo {year} {2014})},\ \Eprint {http://arxiv.org/abs/1306.6870}
  {arXiv:1306.6870 [astro-ph.CO]} \BibitemShut {NoStop}%
\bibitem [{\citenamefont {Schneider}\ \emph {et~al.}(1992)\citenamefont
  {Schneider}, \citenamefont {Ehlers},\ and\ \citenamefont
  {Falco}}]{schneider}%
  \BibitemOpen
  \bibfield  {author} {\bibinfo {author} {\bibfnamefont {P.}~\bibnamefont
  {Schneider}}, \bibinfo {author} {\bibfnamefont {J.}~\bibnamefont {Ehlers}}, \
  and\ \bibinfo {author} {\bibfnamefont {E.}~\bibnamefont {Falco}},\
  }\href@noop {} {\emph {\bibinfo {title} {Gravitational Lenses}}}\ (\bibinfo
  {publisher} {Springer},\ \bibinfo {year} {1992})\BibitemShut {NoStop}%
\bibitem [{\citenamefont {Garcia-Fernandez}\ \emph {et~al.}(2018)\citenamefont
  {Garcia-Fernandez} \emph {et~al.}}]{Garcia-Fernandez:2016oud}%
  \BibitemOpen
  \bibfield  {author} {\bibinfo {author} {\bibfnamefont {M.}~\bibnamefont
  {Garcia-Fernandez}} \emph {et~al.} (\bibinfo {collaboration} {DES}),\ }\href
  {\doibase 10.1093/mnras/sty282} {\bibfield  {journal} {\bibinfo  {journal}
  {Mon. Not. Roy. Astron. Soc.}\ }\textbf {\bibinfo {volume} {476}},\ \bibinfo
  {pages} {1071} (\bibinfo {year} {2018})},\ \Eprint
  {http://arxiv.org/abs/1611.10326} {arXiv:1611.10326 [astro-ph.CO]}
  \BibitemShut {NoStop}%
\bibitem [{\citenamefont {Pritchard}\ and\ \citenamefont
  {Loeb}(2012)}]{Pritchard:2011xb}%
  \BibitemOpen
  \bibfield  {author} {\bibinfo {author} {\bibfnamefont {J.~R.}\ \bibnamefont
  {Pritchard}}\ and\ \bibinfo {author} {\bibfnamefont {A.}~\bibnamefont
  {Loeb}},\ }\href {\doibase 10.1088/0034-4885/75/8/086901} {\bibfield
  {journal} {\bibinfo  {journal} {Rept. Prog. Phys.}\ }\textbf {\bibinfo
  {volume} {75}},\ \bibinfo {pages} {086901} (\bibinfo {year} {2012})},\
  \Eprint {http://arxiv.org/abs/1109.6012} {arXiv:1109.6012 [astro-ph.CO]}
  \BibitemShut {NoStop}%
\bibitem [{\citenamefont {{Battye}}\ \emph {et~al.}(2013)\citenamefont
  {{Battye}}, \citenamefont {{Browne}}, \citenamefont {{Dickinson}},
  \citenamefont {{Heron}}, \citenamefont {{Maffei}},\ and\ \citenamefont
  {{Pourtsidou}}}]{2013MNRAS.434.1239B}%
  \BibitemOpen
  \bibfield  {author} {\bibinfo {author} {\bibfnamefont {R.~A.}\ \bibnamefont
  {{Battye}}}, \bibinfo {author} {\bibfnamefont {I.~W.~A.}\ \bibnamefont
  {{Browne}}}, \bibinfo {author} {\bibfnamefont {C.}~\bibnamefont
  {{Dickinson}}}, \bibinfo {author} {\bibfnamefont {G.}~\bibnamefont
  {{Heron}}}, \bibinfo {author} {\bibfnamefont {B.}~\bibnamefont {{Maffei}}}, \
  and\ \bibinfo {author} {\bibfnamefont {A.}~\bibnamefont {{Pourtsidou}}},\
  }\href {\doibase 10.1093/mnras/stt1082} {\bibfield  {journal} {\bibinfo
  {journal} {Mon. Not. Roy. Astron. Soc.}\ }\textbf {\bibinfo {volume} {434}},\
  \bibinfo {pages} {1239} (\bibinfo {year} {2013})},\ \Eprint
  {http://arxiv.org/abs/1209.0343} {arXiv:1209.0343 [astro-ph.CO]} \BibitemShut
  {NoStop}%
\bibitem [{\citenamefont {Newburgh}\ \emph {et~al.}(2016)\citenamefont
  {Newburgh} \emph {et~al.}}]{Newburgh:2016mwi}%
  \BibitemOpen
  \bibfield  {author} {\bibinfo {author} {\bibfnamefont {L.~B.}\ \bibnamefont
  {Newburgh}} \emph {et~al.},\ }\bibfield  {booktitle} {\emph {\bibinfo
  {booktitle} {{Proceedings, Ground-based and Airborne Telescopes VI:
  Edinburgh, United Kingdom, June 26-July 1, 2016}}},\ }\href {\doibase
  10.1117/12.2234286} {\bibfield  {journal} {\bibinfo  {journal} {Proc. SPIE
  Int. Soc. Opt. Eng.}\ }\textbf {\bibinfo {volume} {9906}},\ \bibinfo {pages}
  {99065X} (\bibinfo {year} {2016})},\ \Eprint
  {http://arxiv.org/abs/1607.02059} {arXiv:1607.02059 [astro-ph.IM]}
  \BibitemShut {NoStop}%
\bibitem [{\citenamefont {Newburgh}\ \emph {et~al.}(2014)\citenamefont
  {Newburgh} \emph {et~al.}}]{Newburgh:2014toa}%
  \BibitemOpen
  \bibfield  {author} {\bibinfo {author} {\bibfnamefont {L.~B.}\ \bibnamefont
  {Newburgh}} \emph {et~al.},\ }\href {\doibase 10.1117/12.2056962} {\bibfield
  {journal} {\bibinfo  {journal} {Proc. SPIE Int. Soc. Opt. Eng.}\ }\textbf
  {\bibinfo {volume} {9145}},\ \bibinfo {pages} {4V} (\bibinfo {year}
  {2014})},\ \Eprint {http://arxiv.org/abs/1406.2267} {arXiv:1406.2267
  [astro-ph.IM]} \BibitemShut {NoStop}%
\bibitem [{\citenamefont {Santos}\ \emph {et~al.}(2017)\citenamefont {Santos}
  \emph {et~al.}}]{Santos:2017qgq}%
  \BibitemOpen
  \bibfield  {author} {\bibinfo {author} {\bibfnamefont {M.~G.}\ \bibnamefont
  {Santos}} \emph {et~al.} (\bibinfo {collaboration} {MeerKLASS}),\ }in\
  \href@noop {} {\emph {\bibinfo {booktitle} {{Proceedings, MeerKAT Science: On
  the Pathway to the SKA (MeerKAT2016): Stellenbosch, South Africa, May 25-27,
  2016}}}}\ (\bibinfo {year} {2017})\ \Eprint {http://arxiv.org/abs/1709.06099}
  {arXiv:1709.06099 [astro-ph.CO]} \BibitemShut {NoStop}%
\bibitem [{\citenamefont {Xu}\ \emph {et~al.}(2015)\citenamefont {Xu},
  \citenamefont {Wang},\ and\ \citenamefont {Chen}}]{Xu:2014bya}%
  \BibitemOpen
  \bibfield  {author} {\bibinfo {author} {\bibfnamefont {Y.}~\bibnamefont
  {Xu}}, \bibinfo {author} {\bibfnamefont {X.}~\bibnamefont {Wang}}, \ and\
  \bibinfo {author} {\bibfnamefont {X.}~\bibnamefont {Chen}},\ }\href {\doibase
  10.1088/0004-637X/798/1/40} {\bibfield  {journal} {\bibinfo  {journal}
  {Astrophys. J.}\ }\textbf {\bibinfo {volume} {798}},\ \bibinfo {pages} {40}
  (\bibinfo {year} {2015})},\ \Eprint {http://arxiv.org/abs/1410.7794}
  {arXiv:1410.7794 [astro-ph.CO]} \BibitemShut {NoStop}%
\bibitem [{\citenamefont {Slosar}\ \emph {et~al.}(2019)\citenamefont {Slosar}
  \emph {et~al.}}]{Bandura:2019uvb}%
  \BibitemOpen
  \bibfield  {author} {\bibinfo {author} {\bibfnamefont {A.}~\bibnamefont
  {Slosar}} \emph {et~al.} (\bibinfo {collaboration} {PUMA}),\ }\href@noop {}
  {\  (\bibinfo {year} {2019})},\ \Eprint {http://arxiv.org/abs/1907.12559}
  {arXiv:1907.12559 [astro-ph.IM]} \BibitemShut {NoStop}%
\bibitem [{\citenamefont {Bull}(2016)}]{Bull:2015lja}%
  \BibitemOpen
  \bibfield  {author} {\bibinfo {author} {\bibfnamefont {P.}~\bibnamefont
  {Bull}},\ }\href {\doibase 10.3847/0004-637X/817/1/26} {\bibfield  {journal}
  {\bibinfo  {journal} {Astrophys. J.}\ }\textbf {\bibinfo {volume} {817}},\
  \bibinfo {pages} {26} (\bibinfo {year} {2016})},\ \Eprint
  {http://arxiv.org/abs/1509.07562} {arXiv:1509.07562 [astro-ph.CO]}
  \BibitemShut {NoStop}%
\bibitem [{\citenamefont {Hall}\ \emph {et~al.}(2013)\citenamefont {Hall},
  \citenamefont {Bonvin},\ and\ \citenamefont {Challinor}}]{Hall:2012wd}%
  \BibitemOpen
  \bibfield  {author} {\bibinfo {author} {\bibfnamefont {A.}~\bibnamefont
  {Hall}}, \bibinfo {author} {\bibfnamefont {C.}~\bibnamefont {Bonvin}}, \ and\
  \bibinfo {author} {\bibfnamefont {A.}~\bibnamefont {Challinor}},\ }\href
  {\doibase 10.1103/PhysRevD.87.064026} {\bibfield  {journal} {\bibinfo
  {journal} {Phys. Rev.}\ }\textbf {\bibinfo {volume} {D87}},\ \bibinfo {pages}
  {064026} (\bibinfo {year} {2013})},\ \Eprint {http://arxiv.org/abs/1212.0728}
  {arXiv:1212.0728 [astro-ph.CO]} \BibitemShut {NoStop}%
\bibitem [{\citenamefont {Jalilvand}\ \emph {et~al.}(2019)\citenamefont
  {Jalilvand}, \citenamefont {Majerotto}, \citenamefont {Durrer},\ and\
  \citenamefont {Kunz}}]{Jalivand:2018vfz}%
  \BibitemOpen
  \bibfield  {author} {\bibinfo {author} {\bibfnamefont {M.}~\bibnamefont
  {Jalilvand}}, \bibinfo {author} {\bibfnamefont {E.}~\bibnamefont
  {Majerotto}}, \bibinfo {author} {\bibfnamefont {R.}~\bibnamefont {Durrer}}, \
  and\ \bibinfo {author} {\bibfnamefont {M.}~\bibnamefont {Kunz}},\ }\href
  {\doibase 10.1088/1475-7516/2019/01/020} {\bibfield  {journal} {\bibinfo
  {journal} {JCAP}\ }\textbf {\bibinfo {volume} {1901}},\ \bibinfo {pages}
  {020} (\bibinfo {year} {2019})},\ \Eprint {http://arxiv.org/abs/1807.01351}
  {arXiv:1807.01351 [astro-ph.CO]} \BibitemShut {NoStop}%
\bibitem [{\citenamefont {Yoo}\ \emph {et~al.}(2009)\citenamefont {Yoo},
  \citenamefont {Fitzpatrick},\ and\ \citenamefont {Zaldarriaga}}]{Yoo:2009au}%
  \BibitemOpen
  \bibfield  {author} {\bibinfo {author} {\bibfnamefont {J.}~\bibnamefont
  {Yoo}}, \bibinfo {author} {\bibfnamefont {A.~L.}\ \bibnamefont
  {Fitzpatrick}}, \ and\ \bibinfo {author} {\bibfnamefont {M.}~\bibnamefont
  {Zaldarriaga}},\ }\href {\doibase 10.1103/PhysRevD.80.083514} {\bibfield
  {journal} {\bibinfo  {journal} {Phys. Rev.}\ }\textbf {\bibinfo {volume}
  {D80}},\ \bibinfo {pages} {083514} (\bibinfo {year} {2009})},\ \Eprint
  {http://arxiv.org/abs/0907.0707} {arXiv:0907.0707 [astro-ph.CO]} \BibitemShut
  {NoStop}%
\bibitem [{\citenamefont {Bonvin}\ and\ \citenamefont
  {Durrer}(2011)}]{Bonvin:2011bg}%
  \BibitemOpen
  \bibfield  {author} {\bibinfo {author} {\bibfnamefont {C.}~\bibnamefont
  {Bonvin}}\ and\ \bibinfo {author} {\bibfnamefont {R.}~\bibnamefont
  {Durrer}},\ }\href {\doibase 10.1103/PhysRevD.84.063505} {\bibfield
  {journal} {\bibinfo  {journal} {Phys. Rev.}\ }\textbf {\bibinfo {volume}
  {D84}},\ \bibinfo {pages} {063505} (\bibinfo {year} {2011})},\ \Eprint
  {http://arxiv.org/abs/1105.5280} {arXiv:1105.5280 [astro-ph.CO]} \BibitemShut
  {NoStop}%
\bibitem [{\citenamefont {Challinor}\ and\ \citenamefont
  {Lewis}(2011)}]{Challinor:2011bk}%
  \BibitemOpen
  \bibfield  {author} {\bibinfo {author} {\bibfnamefont {A.}~\bibnamefont
  {Challinor}}\ and\ \bibinfo {author} {\bibfnamefont {A.}~\bibnamefont
  {Lewis}},\ }\href {\doibase 10.1103/PhysRevD.84.043516} {\bibfield  {journal}
  {\bibinfo  {journal} {Phys. Rev.}\ }\textbf {\bibinfo {volume} {D84}},\
  \bibinfo {pages} {043516} (\bibinfo {year} {2011})},\ \Eprint
  {http://arxiv.org/abs/1105.5292} {arXiv:1105.5292 [astro-ph.CO]} \BibitemShut
  {NoStop}%
\bibitem [{\citenamefont {Jeong}\ \emph {et~al.}(2012)\citenamefont {Jeong},
  \citenamefont {Schmidt},\ and\ \citenamefont {Hirata}}]{Jeong:2011as}%
  \BibitemOpen
  \bibfield  {author} {\bibinfo {author} {\bibfnamefont {D.}~\bibnamefont
  {Jeong}}, \bibinfo {author} {\bibfnamefont {F.}~\bibnamefont {Schmidt}}, \
  and\ \bibinfo {author} {\bibfnamefont {C.~M.}\ \bibnamefont {Hirata}},\
  }\href {\doibase 10.1103/PhysRevD.85.023504} {\bibfield  {journal} {\bibinfo
  {journal} {Phys. Rev.}\ }\textbf {\bibinfo {volume} {D85}},\ \bibinfo {pages}
  {023504} (\bibinfo {year} {2012})},\ \Eprint {http://arxiv.org/abs/1107.5427}
  {arXiv:1107.5427 [astro-ph.CO]} \BibitemShut {NoStop}%
\bibitem [{DES()}]{DESsurvey}%
  \BibitemOpen
  \href@noop {} {}\bibinfo {howpublished}
  {\url{https://www.darkenergysurvey.org/}}\BibitemShut {NoStop}%
\bibitem [{\citenamefont {{Planck Collaboration}}\ \emph
  {et~al.}(2018)\citenamefont {{Planck Collaboration}}, \citenamefont
  {{Aghanim}}, \citenamefont {{Akrami}}, \citenamefont {{Ashdown}},
  \citenamefont {{Aumont}}, \citenamefont {{Baccigalupi}}, \citenamefont
  {{Ballardini}}, \citenamefont {{Banday}}, \citenamefont {{Barreiro}},\ and\
  \citenamefont {{Bartolo}}}]{Aghanim:2018eyx}%
  \BibitemOpen
  \bibfield  {author} {\bibinfo {author} {\bibnamefont {{Planck
  Collaboration}}}, \bibinfo {author} {\bibfnamefont {N.}~\bibnamefont
  {{Aghanim}}}, \bibinfo {author} {\bibfnamefont {Y.}~\bibnamefont {{Akrami}}},
  \bibinfo {author} {\bibfnamefont {M.}~\bibnamefont {{Ashdown}}}, \bibinfo
  {author} {\bibfnamefont {J.}~\bibnamefont {{Aumont}}}, \bibinfo {author}
  {\bibfnamefont {C.}~\bibnamefont {{Baccigalupi}}}, \bibinfo {author}
  {\bibfnamefont {M.}~\bibnamefont {{Ballardini}}}, \bibinfo {author}
  {\bibfnamefont {A.~J.}\ \bibnamefont {{Banday}}}, \bibinfo {author}
  {\bibfnamefont {R.~B.}\ \bibnamefont {{Barreiro}}}, \ and\ \bibinfo {author}
  {\bibfnamefont {N.}~\bibnamefont {{Bartolo}}},\ }\href@noop {} {\bibfield
  {journal} {\bibinfo  {journal} {arXiv e-prints}\ ,\ \bibinfo {eid}
  {arXiv:1807.06209}} (\bibinfo {year} {2018})},\ \Eprint
  {http://arxiv.org/abs/1807.06209} {arXiv:1807.06209 [astro-ph.CO]}
  \BibitemShut {NoStop}%
\bibitem [{\citenamefont {Castorina}\ and\ \citenamefont
  {Villaescusa-Navarro}(2017)}]{Castorina:2016bfm}%
  \BibitemOpen
  \bibfield  {author} {\bibinfo {author} {\bibfnamefont {E.}~\bibnamefont
  {Castorina}}\ and\ \bibinfo {author} {\bibfnamefont {F.}~\bibnamefont
  {Villaescusa-Navarro}},\ }\href {\doibase 10.1093/mnras/stx1599} {\bibfield
  {journal} {\bibinfo  {journal} {Mon. Not. Roy. Astron. Soc.}\ }\textbf
  {\bibinfo {volume} {471}},\ \bibinfo {pages} {1788} (\bibinfo {year}
  {2017})},\ \Eprint {http://arxiv.org/abs/1609.05157} {arXiv:1609.05157
  [astro-ph.CO]} \BibitemShut {NoStop}%
\bibitem [{\citenamefont {Bull}\ \emph {et~al.}(2015)\citenamefont {Bull},
  \citenamefont {Ferreira}, \citenamefont {Patel},\ and\ \citenamefont
  {Santos}}]{Bull_HI_noise}%
  \BibitemOpen
  \bibfield  {author} {\bibinfo {author} {\bibfnamefont {P.}~\bibnamefont
  {Bull}}, \bibinfo {author} {\bibfnamefont {P.~G.}\ \bibnamefont {Ferreira}},
  \bibinfo {author} {\bibfnamefont {P.}~\bibnamefont {Patel}}, \ and\ \bibinfo
  {author} {\bibfnamefont {M.~G.}\ \bibnamefont {Santos}},\ }\href {\doibase
  10.1088/0004-637X/803/1/21} {\bibfield  {journal} {\bibinfo  {journal}
  {Astrophys. J.}\ }\textbf {\bibinfo {volume} {803}},\ \bibinfo {pages} {21}
  (\bibinfo {year} {2015})},\ \Eprint {http://arxiv.org/abs/1405.1452}
  {arXiv:1405.1452 [astro-ph.CO]} \BibitemShut {NoStop}%
\bibitem [{\citenamefont {Zaldarriaga}\ \emph {et~al.}(2004)\citenamefont
  {Zaldarriaga}, \citenamefont {Furlanetto},\ and\ \citenamefont
  {Hernquist}}]{Zaldarriaga:2003du}%
  \BibitemOpen
  \bibfield  {author} {\bibinfo {author} {\bibfnamefont {M.}~\bibnamefont
  {Zaldarriaga}}, \bibinfo {author} {\bibfnamefont {S.~R.}\ \bibnamefont
  {Furlanetto}}, \ and\ \bibinfo {author} {\bibfnamefont {L.}~\bibnamefont
  {Hernquist}},\ }\href {\doibase 10.1086/386327} {\bibfield  {journal}
  {\bibinfo  {journal} {Astrophys. J.}\ }\textbf {\bibinfo {volume} {608}},\
  \bibinfo {pages} {622} (\bibinfo {year} {2004})},\ \Eprint
  {http://arxiv.org/abs/astro-ph/0311514} {arXiv:astro-ph/0311514 [astro-ph]}
  \BibitemShut {NoStop}%
\bibitem [{\citenamefont {Pourtsidou}\ and\ \citenamefont
  {Metcalf}(2014)}]{Pourtsidou:2013hea}%
  \BibitemOpen
  \bibfield  {author} {\bibinfo {author} {\bibfnamefont {A.}~\bibnamefont
  {Pourtsidou}}\ and\ \bibinfo {author} {\bibfnamefont {R.~B.}\ \bibnamefont
  {Metcalf}},\ }\href {\doibase 10.1093/mnrasl/slt175} {\bibfield  {journal}
  {\bibinfo  {journal} {Mon. Not. Roy. Astron. Soc.}\ }\textbf {\bibinfo
  {volume} {439}},\ \bibinfo {pages} {L36} (\bibinfo {year} {2014})},\ \Eprint
  {http://arxiv.org/abs/1311.4484} {arXiv:1311.4484 [astro-ph.CO]} \BibitemShut
  {NoStop}%
\bibitem [{\citenamefont {Shaw}\ \emph {et~al.}(2014)\citenamefont {Shaw},
  \citenamefont {Sigurdson}, \citenamefont {Pen}, \citenamefont {Stebbins},\
  and\ \citenamefont {Sitwell}}]{Shaw:2013wza}%
  \BibitemOpen
  \bibfield  {author} {\bibinfo {author} {\bibfnamefont {J.~R.}\ \bibnamefont
  {Shaw}}, \bibinfo {author} {\bibfnamefont {K.}~\bibnamefont {Sigurdson}},
  \bibinfo {author} {\bibfnamefont {U.-L.}\ \bibnamefont {Pen}}, \bibinfo
  {author} {\bibfnamefont {A.}~\bibnamefont {Stebbins}}, \ and\ \bibinfo
  {author} {\bibfnamefont {M.}~\bibnamefont {Sitwell}},\ }\href {\doibase
  10.1088/0004-637X/781/2/57} {\bibfield  {journal} {\bibinfo  {journal}
  {Astrophys. J.}\ }\textbf {\bibinfo {volume} {781}},\ \bibinfo {pages} {57}
  (\bibinfo {year} {2014})},\ \Eprint {http://arxiv.org/abs/1302.0327}
  {arXiv:1302.0327 [astro-ph.CO]} \BibitemShut {NoStop}%
\bibitem [{\citenamefont {Shaw}\ \emph {et~al.}(2015)\citenamefont {Shaw},
  \citenamefont {Sigurdson}, \citenamefont {Sitwell}, \citenamefont
  {Stebbins},\ and\ \citenamefont {Pen}}]{Shaw:2014khi}%
  \BibitemOpen
  \bibfield  {author} {\bibinfo {author} {\bibfnamefont {J.~R.}\ \bibnamefont
  {Shaw}}, \bibinfo {author} {\bibfnamefont {K.}~\bibnamefont {Sigurdson}},
  \bibinfo {author} {\bibfnamefont {M.}~\bibnamefont {Sitwell}}, \bibinfo
  {author} {\bibfnamefont {A.}~\bibnamefont {Stebbins}}, \ and\ \bibinfo
  {author} {\bibfnamefont {U.-L.}\ \bibnamefont {Pen}},\ }\href {\doibase
  10.1103/PhysRevD.91.083514} {\bibfield  {journal} {\bibinfo  {journal} {Phys.
  Rev.}\ }\textbf {\bibinfo {volume} {D91}},\ \bibinfo {pages} {083514}
  (\bibinfo {year} {2015})},\ \Eprint {http://arxiv.org/abs/1401.2095}
  {arXiv:1401.2095 [astro-ph.CO]} \BibitemShut {NoStop}%
\bibitem [{\citenamefont {Font-Ribera}\ \emph {et~al.}(2014)\citenamefont
  {Font-Ribera}, \citenamefont {McDonald}, \citenamefont {Mostek},
  \citenamefont {Reid}, \citenamefont {Seo},\ and\ \citenamefont
  {Slosar}}]{Font-Ribera:2013rwa}%
  \BibitemOpen
  \bibfield  {author} {\bibinfo {author} {\bibfnamefont {A.}~\bibnamefont
  {Font-Ribera}}, \bibinfo {author} {\bibfnamefont {P.}~\bibnamefont
  {McDonald}}, \bibinfo {author} {\bibfnamefont {N.}~\bibnamefont {Mostek}},
  \bibinfo {author} {\bibfnamefont {B.~A.}\ \bibnamefont {Reid}}, \bibinfo
  {author} {\bibfnamefont {H.-J.}\ \bibnamefont {Seo}}, \ and\ \bibinfo
  {author} {\bibfnamefont {A.}~\bibnamefont {Slosar}},\ }\href {\doibase
  10.1088/1475-7516/2014/05/023} {\bibfield  {journal} {\bibinfo  {journal}
  {JCAP}\ }\textbf {\bibinfo {volume} {1405}},\ \bibinfo {pages} {023}
  (\bibinfo {year} {2014})},\ \Eprint {http://arxiv.org/abs/1308.4164}
  {arXiv:1308.4164 [astro-ph.CO]} \BibitemShut {NoStop}%
\bibitem [{\citenamefont {{Ho}}\ \emph {et~al.}(2012)\citenamefont {{Ho}},
  \citenamefont {{Cuesta}}, \citenamefont {{Seo}}, \citenamefont {{de Putter}},
  \citenamefont {{Ross}}, \citenamefont {{White}}, \citenamefont
  {{Padmanabhan}}, \citenamefont {{Saito}},\ and\ \citenamefont
  {et~al.}}]{Ho2012}%
  \BibitemOpen
  \bibfield  {author} {\bibinfo {author} {\bibfnamefont {S.}~\bibnamefont
  {{Ho}}}, \bibinfo {author} {\bibfnamefont {A.}~\bibnamefont {{Cuesta}}},
  \bibinfo {author} {\bibfnamefont {H.-J.}\ \bibnamefont {{Seo}}}, \bibinfo
  {author} {\bibfnamefont {R.}~\bibnamefont {{de Putter}}}, \bibinfo {author}
  {\bibfnamefont {A.~J.}\ \bibnamefont {{Ross}}}, \bibinfo {author}
  {\bibfnamefont {M.}~\bibnamefont {{White}}}, \bibinfo {author} {\bibfnamefont
  {N.}~\bibnamefont {{Padmanabhan}}}, \bibinfo {author} {\bibfnamefont
  {S.}~\bibnamefont {{Saito}}}, \ and\ \bibinfo {author} {\bibnamefont
  {et~al.}},\ }\href {\doibase 10.1088/0004-637X/761/1/14} {\bibfield
  {journal} {\bibinfo  {journal} {\apj}\ }\textbf {\bibinfo {volume} {761}},\
  \bibinfo {eid} {14} (\bibinfo {year} {2012})},\ \Eprint
  {http://arxiv.org/abs/1201.2137} {arXiv:1201.2137 [astro-ph.CO]} \BibitemShut
  {NoStop}%
\bibitem [{\citenamefont {{DES Collaboration}}\ \emph
  {et~al.}(2019)\citenamefont {{DES Collaboration}}, \citenamefont {{Camacho}},
  \citenamefont {{Kokron}}, \citenamefont {{Andrade-Oliveira}}, \citenamefont
  {{Rosenfeld}}, \citenamefont {{Lima}}, \citenamefont {{Lacasa}},
  \citenamefont {{Sobreira}}, \citenamefont {{da Costa}},\ and\ \citenamefont
  {et~al.}}]{Camacho2019}%
  \BibitemOpen
  \bibfield  {author} {\bibinfo {author} {\bibnamefont {{DES Collaboration}}},
  \bibinfo {author} {\bibfnamefont {H.}~\bibnamefont {{Camacho}}}, \bibinfo
  {author} {\bibfnamefont {N.}~\bibnamefont {{Kokron}}}, \bibinfo {author}
  {\bibfnamefont {F.}~\bibnamefont {{Andrade-Oliveira}}}, \bibinfo {author}
  {\bibfnamefont {R.}~\bibnamefont {{Rosenfeld}}}, \bibinfo {author}
  {\bibfnamefont {M.}~\bibnamefont {{Lima}}}, \bibinfo {author} {\bibfnamefont
  {F.}~\bibnamefont {{Lacasa}}}, \bibinfo {author} {\bibfnamefont
  {F.}~\bibnamefont {{Sobreira}}}, \bibinfo {author} {\bibfnamefont {L.~N.}\
  \bibnamefont {{da Costa}}}, \ and\ \bibinfo {author} {\bibnamefont
  {et~al.}},\ }\href {\doibase 10.1093/mnras/stz1514} {\bibfield  {journal}
  {\bibinfo  {journal} {\mnras}\ }\textbf {\bibinfo {volume} {487}},\ \bibinfo
  {pages} {3870} (\bibinfo {year} {2019})},\ \Eprint
  {http://arxiv.org/abs/1807.10163} {arXiv:1807.10163 [astro-ph.CO]}
  \BibitemShut {NoStop}%
\bibitem [{\citenamefont {{DES Collaboration}}\ \emph
  {et~al.}(2016)\citenamefont {{DES Collaboration}}, \citenamefont {{Crocce}}
  \emph {et~al.}}]{Crocce2016}%
  \BibitemOpen
  \bibfield  {author} {\bibinfo {author} {\bibnamefont {{DES Collaboration}}},
  \bibinfo {author} {\bibfnamefont {M.}~\bibnamefont {{Crocce}}},  \emph
  {et~al.},\ }\href {\doibase 10.1093/mnras/stv2590} {\bibfield  {journal}
  {\bibinfo  {journal} {Mon. Not. Roy. Astron. Soc.}\ }\textbf {\bibinfo
  {volume} {455}},\ \bibinfo {pages} {4301} (\bibinfo {year} {2016})},\ \Eprint
  {http://arxiv.org/abs/1507.05360} {arXiv:1507.05360 [astro-ph.CO]}
  \BibitemShut {NoStop}%
\bibitem [{\citenamefont {{DES Collaboration}}\ \emph
  {et~al.}(2018)\citenamefont {{DES Collaboration}}, \citenamefont
  {{Elvin-Poole}} \emph {et~al.}}]{Elvin-Poole2018}%
  \BibitemOpen
  \bibfield  {author} {\bibinfo {author} {\bibnamefont {{DES Collaboration}}},
  \bibinfo {author} {\bibfnamefont {J.}~\bibnamefont {{Elvin-Poole}}},  \emph
  {et~al.},\ }\href {\doibase 10.1103/PhysRevD.98.042006} {\bibfield  {journal}
  {\bibinfo  {journal} {\prd}\ }\textbf {\bibinfo {volume} {98}},\ \bibinfo
  {eid} {042006} (\bibinfo {year} {2018})},\ \Eprint
  {http://arxiv.org/abs/1708.01536} {arXiv:1708.01536 [astro-ph.CO]}
  \BibitemShut {NoStop}%
\bibitem [{\citenamefont {Montanari}\ and\ \citenamefont
  {Durrer}(2015)}]{Montanari:2015rga}%
  \BibitemOpen
  \bibfield  {author} {\bibinfo {author} {\bibfnamefont {F.}~\bibnamefont
  {Montanari}}\ and\ \bibinfo {author} {\bibfnamefont {R.}~\bibnamefont
  {Durrer}},\ }\href {\doibase 10.1088/1475-7516/2015/10/070} {\bibfield
  {journal} {\bibinfo  {journal} {JCAP}\ }\textbf {\bibinfo {volume} {1510}},\
  \bibinfo {pages} {070} (\bibinfo {year} {2015})},\ \Eprint
  {http://arxiv.org/abs/1506.01369} {arXiv:1506.01369 [astro-ph.CO]}
  \BibitemShut {NoStop}%
\bibitem [{\citenamefont {Bull}(2015)}]{radiofisher}%
  \BibitemOpen
  \bibfield  {author} {\bibinfo {author} {\bibfnamefont {P.}~\bibnamefont
  {Bull}},\ }\href@noop {} {\enquote {\bibinfo {title} {Radiofisher},}\
  }\bibinfo {howpublished}
  {\url{https://gitlab.com/radio-fisher/bao21cm/tree/master/radiofisher}}
  (\bibinfo {year} {2015})\BibitemShut {NoStop}%
\end{thebibliography}%

\end{document}